\documentclass[twocolumn,showpacs,preprintnumbers,final,a4paper,aps,citeautoscript,footinbib,pra,superscriptaddress]{revtex4-2}
\usepackage{amsmath,amssymb,amsfonts}
\usepackage{braket}
\usepackage[pdftex]{graphicx,color}
\graphicspath{{./figures/}}
\usepackage[latin1]{inputenc}
\usepackage{times}
\usepackage{mathrsfs}
\usepackage[normalem]{ulem}
\usepackage{empheq}
\usepackage[sort&compress]{natbib}
\usepackage[colorlinks,bookmarks=true,citecolor=blue,linkcolor=blue,urlcolor=blue,breaklinks=true]{hyperref}
\usepackage{simplewick}
\usepackage[latin1]{inputenc}
\usepackage[vcentermath]{youngtab}
\usepackage{theorem}

\newcommand{\be}{\text{e}}

\begin{document}
\title{Ferromagnetism in the SU($N$) Kondo lattice model  -- SU($N$) double exchange and supersymmetry
}
\author{Keisuke Totsuka}
\affiliation{Center for Gravitational Physics and Quantum Information, 
Yukawa Institute for Theoretical Physics, Kyoto University, Kyoto 606-8502, Japan}
\date{\today}
\begin{abstract}
We study the ground-state properties of the SU($N$)-generalization of the Kondo-lattice model in one dimension 
when the Kondo coupling $J_{\text{K}}$ (both ferromagnetic and antiferromagnetic) is sufficiently strong.  
Both cases can be realized using alkaline-earth-metal-like cold gases in optical lattices.  
Specifically, we first carry out the strong-coupling expansion and identify two insulating phases (one of which is 
the SU($N$)-analog of the well-known gapped Kondo singlet phase).  We then rigorously establish that 
the ground state in the low-density (for $J_{\text{K}}<0$) or the high-density (for $J_{\text{K}}>0$) region 
is ferromagnetic.  The results are accounted for by generalizing the double-exchange mechanism to SU($N$) ``spins''.  
Possible realizations of Bose-Fermi supersymmetry SU($N | 1$) in the (generalized) SU($N$) Kondo-lattice model 
are discussed as well.  
\end{abstract}
\maketitle
\section{Introduction}
\label{sec:intro}
In physics, high symmetries have been playing crucial roles in a unified understanding of seemingly different phenomena.  
In such situations, we often work with simple unifying theories based on high symmetries and 
try to understand the actual phases by taking into account the deviation from the idealized high symmetries.  
In condensed-matter physics, systems with high SU($N$)-symmetry ($N \geq 3$) have been studied for a few decades and 
a variety of intriguing properties have been predicted so far.   However, in the standard solid-state settings, 
the realization of SU($N$) symmetry exploits, on top of the spin-SU(2), 
additional symmetries [e.g., SU(2)-symmetry associated with 
orbital, valley, multiple layers, etc.] that necessitate some sort of fine-tuning or idealization 
\cite{Li-M-S-Z-98,Wu-S-A-M-J-14}.   So far, it has not been so clear 
to what extent physics found in those idealized systems with perfect SU($N$) symmetry is robust against possible deviations 
in realistic systems.   
The situation changed when the possibility of realizing systems with almost perfect SU($N$)-symmetry using 
alkaline-earth(-like) cold gases has been recognized \cite{Cazalilla-H-U-09,Gorshkov-et-al-10}.   
This has paved the way for testing a variety of remarkable predictions made in SU($N$) fermion and spin systems 
\cite{Cazalilla-R-14,Capponi-L-T-16} in clean and well-controlled settings.  
For instance, the SU($N$) Mott insulator has been realized experimentally \cite{Taie-Y-S-T-12,Hofrichter-SUN-Mott-16},  
in which antiferromagnetic correlation among the localized SU($N$) magnetic moments 
has been observed \cite{Ozawa-etal-double-well-18,Taie-et-al-SUN-AF-22}.  
These are the first steps toward the quantum simulation of even more exotic states of matter, 
e.g., SU($N$) quantum spin liquids \cite{Hermele-G-11,Corboz-L-L-P-M-12,Chen-et-al-SUN-CSL-21}.  

The SU($N$)-symmetric cold gases also provide us with a playground for multi-component itinerant  
fermions that are predicted to exhibit a variety of interesting phenomena 
such as the color superfluidity \cite{Honerkamp-H-04,Cherng-R-D-07}, 
the ``baryonic'' multiple-fermion bound states (dubbed trion when $N=3$) 
\cite{Lecheminant-B-A-05,Rapp-Z-H-H-07,Capponi-R-L-A-B-W-08}, 
the generalized $\eta$-pairing \cite{Yoshida-K-22},  
and itinerant ferromagnetism \cite{Cazalilla-H-U-09,Katsura-T-13,Yip-H-K-14,Li-L-W-14,Bobrow-S-L-18,Tamura-K-21}.  
One of the merits of using the alkaline-earth-metal fermions is that one can easily implement two additional 
``orbital'' degrees of freedom [associated with the two SU($N$)-symmetric atomic states $g$ and $e$] 
that enable us to simulate Kondo physics.  
Since the suggestion of exploring the heavy-fermion physics with the two-orbital alkaline-earth-metal fermions 
\cite{Gorshkov-et-al-10}, some advances have been made both theoretically \cite{Foss-Feig-H-R-10,Nakagawa-K-15,Isaev-R-15} 
and experimentally \cite{Riegger-D-H-F-B-F-18,Ono-A-H-S-T-21}.  
Although most cold-atom literature focuses on the Kondo or heavy-fermion physics in which the local moments 
tend to be screened by the itinerant fermions, there is yet another important state of matter, itinerant ferromagnetism, 
in the Kondo lattice model.   
In fact, in the usual ($N=2$) Kondo lattice model, it is known that the so-called double-exchange mechanism 
\cite{Zener-51,Anderson-H-55,deGennes-60},  
which has been originally introduced in the context of the manganites, stabilizes ferromagnetism when the Kondo coupling 
is ferromagnetically large \cite{Kubo-82,Yunoki-H-M-M-F-D-98,Dagotto-Y-M-M-H-C-P-F-98}, 
and even when it is anti-ferromagnetic, metallic ferromagnetism is favored 
for sufficiently large Kondo coupling \cite{Sigrist-T-U-R-92,Troyer-W-93,McCulloch-J-R-G-02,Peters-K-12} 
(see, e.g., Refs.~\cite{Tsunetsugu-S-U-97,Gulacsi-review-04} for reviews of the one-dimensional Kondo-lattice model).  

In this paper, we will rigorously show that, even for $N \geq 3$, ferromagnetism is one of the dominant phases in 
the one-dimensional SU($N$) Kondo lattice model.  
Some rigorous results have been obtained so far on the SU($N$) itinerant 
ferromagnetism \cite{Katsura-T-13,Li-L-W-14,Bobrow-S-L-18,Tamura-K-21}.  
What we will establish here occurs in a relatively simple setting and 
for a wide range of fermion densities, and only needs relatively loose conditions.  
The SU($N$) Kondo-lattice(-type) systems not only exhibit ferromagnetic and other interesting phases 
but also provide us with a natural playground for emergent Bose-Fermi supersymmetry.  
We will try to demonstrate how supersymmetry is implemented into the low-energy degrees of freedom 
of the Kondo-lattice model.  

This paper is structured as follows.  
In Sec.~\ref{sec:model}, we introduce the SU($N$) Kondo lattice model as a particular limit of  
the two-orbital SU($N$) Hubbard model, which can be realized using alkaline-earth-metal cold Fermi gases.  
Then, we will quickly discuss the symmetries of the model which are useful in capturing the global phase structure.   
Section \ref{sec:strong-coupling} is devoted to the determination of the ground-state phases for 
strong Kondo coupling.  After identifying the ground state in the strong-coupling limit, we derive the low-energy 
effective Hamiltonians both for ferromagnetic and antiferromagnetic Kondo coupling by taking into account 
the lowest-order corrections from the kinetic energy.  
We use these results to prove ferromagnetic ground states in Sec.~\ref{sec:ferromagnetism}.  When $J_{\text{K}}<0$, 
the effective Hamiltonian satisfies the conditions of the Perron-Frobenius theorem, and ferromagnetism in the ground state 
follows immediately.  When $J_{\text{K}}$ is antiferromagnetic, on the other hand, 
the lowest-order effective Hamiltonian does not resolve the huge SU($N$) ``spin'' degeneracy.  
We show that taking into account higher-order 
corrections lift the degeneracy thereby stabilizing ferromagnetism.  
We also explain ferromagnetism in the ferromagnetic ($J_{\text{K}}<0$) Kondo-lattice model by generalizing the double-exchange 
mechanism to SU($N$).  These are the central results of this paper. 

The emergent Bose-Fermi supersymmetry SU($N|1$) in the SU($N$) Kondo-Heisenberg model, which is a variant of 
the SU($N$) Kondo-lattice model, will be discussed in Sec.~\ref{sec:SUSY}.  
The main results and some technical details are summarized in Sec.~\ref{sec:conclusion} and in the Appendixes, 
respectively.  
\section{Model}
\label{sec:model}
\subsection{Two-orbital Hubbard model for alkaline-earth-metal cold fermions}
\label{sec:two-orbital-Hubbard}
To obtain the SU($N$)-symmetric Kondo lattice model, we start from 
the minimal model that describes the alkaline-earth-metal cold fermions loaded in an optical lattice \cite{Gorshkov-et-al-10}: 
\begin{equation}
\begin{split}
&  \mathcal{H}_{\text{G}}  \\
&=  -  \sum_{i}\sum_{m=g,e} t^{(m)} \sum_{\alpha=1}^{N} 
   \left(c_{m\alpha,\,i}^\dag c_{m\alpha,\,i+1}  + \text{H.c.}\right)
 \\
   & +\sum_{i}\sum_{m=g,e} \frac{U^{(m)}}{2} n_{m,\,i}(n_{m,\,i} - 1)  
 -  \sum_{i}\sum_{m=g,e} \mu^{(m)}_{i} n_{m,\,i}  \\
&  + V_{\text{H}}^{g\text{-}e} \sum_i n_{g,\,i} n_{e,\,i} + V_{\text{ex}}^{g\text{-}e} \sum_{i,\alpha \beta} 
  c_{g\alpha,\,i}^\dag c_{e\beta,\,i}^\dag 
  c_{g\beta ,\,i} c_{e\alpha,\,i}   \;  ,
  \end{split}
\label{eqn:2-orbital-Hubbard}
\end{equation} 
where $c_{m\alpha,\,i}^{\dagger}$ ($c_{m\alpha,\,i}$) creates (annihilates) a fermion 
of the color $\alpha\,(=1,\ldots,N)$ in the ``orbital''-$m\,(=g,e)$ at site-$i$, 
and $n_{m,\,i}$ is the corresponding number operator $n_{m,\,i}=\sum_{\alpha} c_{m\alpha,\,i}^{\dagger}c_{m\alpha,\,i}$.  
In alkaline-earth-metal cold fermions, the orbital $g$ ($e$) corresponds to the atomic state ${}^{1} S_{0}$ 
(${}^{3}P_{0}$).   
The local potential $\mu^{(m)}_{i}$ can be site-dependent in general (especially in the cold-atom setting).  
The interactions $U^{(m)}$ and $V_{\text{H}}$ respectively are the Hubbard interactions among the same species 
of fermions and the density-density interaction between the $g$ and $e$ fermions. 
The last term is the exchange interactions between the fermions in different orbitals, which can be 
conveniently recast as:
\begin{equation}
- V_{\text{ex}}^{g\text{-}e}
\sum_{i}\left( 
\sum_{A=1}^{N^{2}-1}\hat{S}_{g,i}^{A}\hat{S}_{e,i}^{A} 
\right) 
- \frac{1}{N} V_{\text{ex}}^{g\text{-}e} \sum_{i}n_{g,i}n_{e,i}
\end{equation}
with $\hat{S}_{g,i}^{A}$ and $\hat{S}_{e,i}^{A}$ being  
the second-quantized SU($N$) spins for the $g$ and $e$ fermions, respectively:
\begin{equation}
\hat{S}_{m,i}^{A} := \sum_{\alpha,\beta=1}^{N} c^{\dagger}_{m\alpha,i} [G^{A}]_{\alpha\beta} \, c_{m\beta,i}  
\quad (m=g,e) \;  .
\label{eqn:second-quantized-gen}
\end{equation}
The $N$-dimensional matrices $G^{A}$ ($A=1,\ldots, N^{2}-1$) are the SU($N$) generators normalized as 
$\text{Tr}(G^{A}G^{B})=\delta^{AB}$, and satisfy:
\begin{equation}
\begin{split}
& [ G^{A} , G^{B} ] = i f^{ABC} G^{C}  \\
& \sum_{A} [G^{A}]_{\alpha\beta} [G^{A}]_{\mu\nu} 
= \delta_{\alpha\nu} \delta_{\beta\mu} - \delta_{\alpha\beta}\delta_{\mu\nu} /N  \\
& (\alpha,\beta,\mu, \nu =1, \ldots, N) \; .
\end{split}
\end{equation}
Basically, the exchange interaction $V_{\text{ex}}^{g\text{-}e}$ is the same as the Hund coupling which is ferromagnetic  
except that here it comes from the atom-atom collision and can be both ferromagnetic and antiferromagnetic. 
Note that there is no hybridization between the $g$ and $e$ fermions which may potentially lead to 
the mixed-valence physics in heavy-fermion systems \cite{Coleman-book-15}.  

\begin{figure}[htb]
\begin{center}
\includegraphics[scale=0.6]{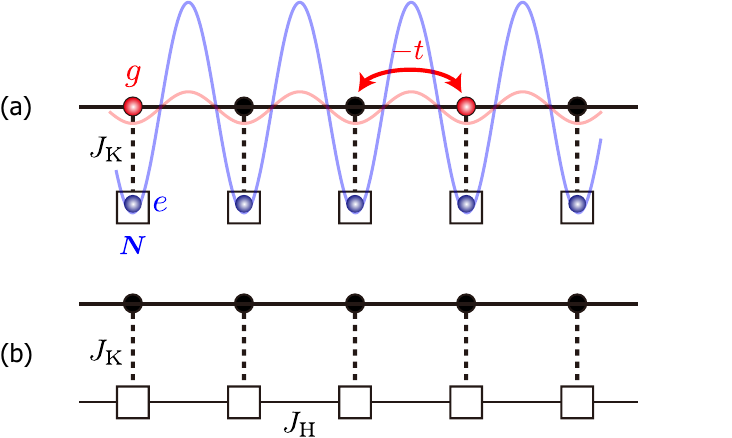}
\end{center}
\caption{(a) The SU($N$) Kondo lattice model \eqref{eqn:SUN-KLM-2} 
with the SU($N$) local moments in $\boldsymbol{N}$ (i.e., the $N$-dimensional 
defining representation \raisebox{-0.2ex}{$\square$}) and its realization with a state-dependent lattice 
(shown by the red and blue curves).  The $e$-fermions localized at the bottoms of the deeper lattice play 
the role of the local moments. 
(b) The Kondo-Heisenberg model \eqref{eqn:SUN-KHM} with additional interaction $J_{\text{H}}$ 
among the local moments. 
\label{fig:SUN-KLM}}
\end{figure}

Now we turn off the hopping of the $e$ fermions: $t^{(e)}=0$ while keeping $t^{(g)}$ finite.   
Experimentally, this is achieved, e.g., by employing 
the so-called state-dependent lattice (SDL) \cite{Gorshkov-et-al-10,Riegger-D-H-F-B-F-18,Ono-A-H-S-T-21} 
in which the $g$ fermions moving in a shallow lattice remain itinerant while the $e$ fermions are localized 
in a deeper lattice [see Fig.~\ref{fig:SUN-KLM}(a)].   
The model parameters can be estimated for actual optical lattices, e.g., for ${}^{173}\text{Yb}$ and 
the setting used in Ref.~\cite{Ono-A-H-S-T-21} as: 
$t^{(e)}/V_{\text{ex}}^{g\text{-}e} \sim 10^{-2}$, $t^{(e)}/t^{(g)} \sim 10^{-3}$, 
$V_{\text{H}}^{g\text{-}e}  \simeq  V_{\text{ex}}^{g\text{-}e}\,(>0)$, 
$U^{(g)}/V_{\text{ex}}^{g\text{-}e} \sim 10^{-1}$, $U^{(e)} \simeq V_{\text{ex}}^{g\text{-}e}$, 
which suggest that we may treat the $e$-fermions as localized.  

When the deeper lattice sites are uniformly occupied by the $e$-fermions, i.e., $n_{e,i} = n_{e}$ ($=1,\ldots, N$), 
the following $\frac{N!}{n_{e}! (N-n_{e})!}$-plet ``spin'' is formed at each site \footnote{%
To realize the situation $n_{e}=1$, strong enough $U^{(e)}$ is necessary.  As $t^{(e)}$ is negligibly small, we can easily find  
the condition for $U^{(e)}$ with the effects of the harmonic trap (the trap frequency $\omega_{\text{trap}}$) 
taken into account: $U^{(e)} > (1/8) \mathcal{N}_{e}^{2} m \omega^{2}_{\text{trap}} a_{0}^{2}$ ($a_{0}$ 
is the lattice constant  
and $\mathcal{N}_{e}$ denotes the number of the $e$ fermions). This condition is fulfilled in the usual experimental settings.}  
\begin{equation*}
 |{}_{[ \alpha_{1}, \cdots, \alpha_{n_{e} } ] } \rangle 
= c_{\alpha_{1},i}^{\dagger} \cdots c_{\alpha_{n_{e}},i}^{\dagger} |0 \rangle  
\end{equation*}
(the bracket $[\cdots]$ denotes anti-symmetrization).  
By the Fermi statistics, these SU($N$) spin states are anti-symmetric in the spin labels $\{ \alpha_{k} \}$, and 
we denote the corresponding irreducible representation (rank-$n_{e}$ anti-symmetric tensor) by the following Young diagram:
\begin{equation}
\text{\scriptsize $n_{e}$} \left\{ 
{\tiny \yng(1,1,1,1)  }
\right.    \; .
\label{eqn:irrep-local-moment}
\end{equation}
In what follows, we will frequently use similar Young diagrams as the substitute for 
the spin ``$S$'' to specify the SU($N$)-spins (irreducible representations, precisely). 
For a quick explanation of the Young diagrams and the irreducible representations, see Appendix~\ref{sec:Young-diag} 
(for more details and other useful knowledge of SU($N$), see, e.g., Ref.~\cite{Georgi-book-99}).  

Under the condition $n_{e,i} = n_{e}\,(=\text{const.})$, 
the interaction $V_{\text{H}}^{g\text{-}e} \sum_{i} n_{g,\,i} n_{e,\,i}$ in \eqref{eqn:2-orbital-Hubbard} reduces 
to $\widetilde{V}_{\text{H}}^{g\text{-}e} \sum_i n_{g,\,i} n_{e,\,i}  \to n_{e} \widetilde{V}_{\text{H}}^{g\text{-}e} \sum_i n_{g,\,i} $ 
($\widetilde{V}_{\text{H}}^{g\text{-}e} := V_{\text{H}}^{g\text{-}e} - V_{\text{ex}}^{g\text{-}e}/N$ 
\footnote{In typical experimental settings, $\tilde{V}_{\text{H}}^{g\text{-}e} >0$ 
for all $N$ and alkaline-earth fermions. Due to positive $U^{(e)}$ and $\tilde{V}_{\text{H}}^{g\text{-}e}$, 
states with $e$-fermions uniformly occupying the lattice $n_{e,i}=1$ are favored.}), 
which just renormalizes the chemical potential, and can be regarded as a constant in a sector with a fixed fermion number.  
If we drop the Hubbard interaction $U^{(g)}$ temporarily, we obtain the following SU($N$) Kondo lattice model (KLM):
\begin{equation}
\begin{split}
& \mathcal{H}_{\text{KLM}} \\ 
& =   - t \sum_{i} \sum_{\alpha=1}^{N} 
   \left( c_{\alpha,\,i}^\dag c^{\alpha}_{i+1}  + \text{H.c.}\right)   
+ J_{\text{K}}  \sum_{i}\left( 
\sum_{A=1}^{N^{2}-1}\hat{s}_{i}^{A} S_{i}^{A}
\right) 
  \; ,
\end{split}
\label{eqn:SUN-KLM-2}
\end{equation} 
where we have simplified the notations as:
\begin{equation}
\begin{split}
& c_{\alpha,\,i}^\dag = c_{g\alpha,\,i}^\dag \, , \quad  c^{\alpha}_{i} = c_{g\alpha ,\,i} \, , \quad 
\hat{s}_{i}^{A} = \hat{S}_{g,i}^{A} \, , \quad S_{i}^{A}  = \hat{S}_{e,i}^{A}  \\
&  t = t^{(g)},  \quad  J_{\text{K}} = - V_{\text{ex}}^{g\text{-}e}    
\end{split}
\end{equation}
(we shall discuss the effects of the neglected $U^{(g)}$ in Sec.~\ref{sec:effect-residual-int}).  
Throughout this paper, the number of lattice sites is denoted by $L$, 
and we use $\mathcal{N}_{\text{c}}$ and $n_{\text{c}}$ for the total fermion number 
$\sum_{i, \alpha} c_{\alpha,i}^{\dagger} c_{i}^{\alpha} = \sum_{i} n_{i}$ 
and the fermion density (or, the average fermion number per site $\mathcal{N}_{\text{c}}/L$), respectively.  
The fermion filling $f$ ($0 \leq f \leq 1$) is defined by: $f= n_{\text{c}} / N$.  
Also, unless otherwise stated, we consider only the one-dimensional system with an open boundary condition.  

The SU($N$) generalization of the KLM has been originally introduced in the context of 
spin-orbit-coupled heavy-fermion materials in which $N$ is the number of the ground-state $j$-multiplet 
($N=2j+1$) and investigated mostly in the large-$N$ limit \cite{Coleman-83,Read-N-D-84} 
(for recent finite-$N$ studies, see, e.g., Ref.~\cite{Raczkowski-A-20} and references cited therein).   

Although the ``SU($N$) spin'' (the irreducible representation, precisely) of the local moment $S_{i}^{A}$ 
in \eqref{eqn:SUN-KLM-2} is constrained by the fermion statistics to those given by  
\eqref{eqn:irrep-local-moment} in the cold-atom setting, we can think of arbitrary representations $\mathcal{R}$ in principle.   
In what follows, we only consider the case $n_{e}=1$ (exactly one $e$ atom) at each site, i.e., the local moments in 
the $N$-dimensional representation $\mathcal{R} = {\tiny \yng(1)}$ [which is the SU($N$)-counterpart 
of $S=1/2$], as it is the simplest and most realistic 
in the above cold-gas setting.  Note that the situation assumed here is very different from that in the standard large-$N$ 
treatment \cite{Coleman-83,Read-N-D-84} in which $n_{e}$ is proportional to $N$.  

In the heavy-fermion setting, $J_{\text{K}}$ comes from the second-order perturbation in the hybridization between 
the conduction electron and the localized $f$-electron and is bound to be positive and small \cite{Sinjukow-N-02}.  
In this sense, the large-$J_{\text{K}}$ physics we will consider below is hard to access directly in heavy-fermion systems. 
On the other hand, if \eqref{eqn:SUN-KLM-2} is realized 
in cold gases, the sign of $J_{\text{K}} (= - V_{\text{ex}}^{g\text{-}e})$ 
depends on that of the scattering lengths for the $g$-$e$ collision \cite{Gorshkov-et-al-10} and can take both signs, 
thereby allowing us to explore both ferromagnetic and antiferromagnetic SU($N$) KLM at strong coupling.   
In fact, it is known experimentally that $J_{\text{K}}$ is negative (ferromagnetic) for ${}^{87}\text{Sr}$ \cite{Zhang-et-al-Sr-14} 
and ${}^{173}\text{Yb}$ \cite{Cappellini-et-al-14,Scazza-et-al-14}, while it is positive (anti-ferromagnetic) 
for ${}^{171}\text{Yb}$ \cite{Ono-K-A-S-T-19}.  

We may also add an exchange interaction $J_{\text{H}} \,(>0)$ between the neighboring local spins, which can arise, e.g., from 
the virtual hopping ($\sim {t^{(e)}}^{2} / U^{(e)}$) of the almost localized $e$ fermions, to define 
the SU($N$) Kondo-Heisenberg model: 
\begin{equation}
\begin{split}
& \mathcal{H}_{\text{KHM}} \\
& =   - t \sum_{i} \sum_{\alpha=1}^{N} 
   \left(c_{\alpha,\,i}^\dag c_{\alpha,\,i+1}  + \text{H.c.}\right)   
+ J_{\text{K}}  \sum_{i}\left( 
\sum_{A=1}^{N^{2}-1}\hat{s}_{i}^{A} S_{i}^{A}
\right)  \\
& \phantom{=} 
+ J_{\text{H}}  \sum_{i}\left( 
\sum_{A=1}^{N^{2}-1} S_{i}^{A} S_{i+1}^{A}
\right) 
+ V \sum_{i} n_{i} n_{i+1}
  \; .
\end{split}  
\label{eqn:SUN-KHM}
\end{equation} 
In the above, we also have added the density-density interaction $V$ which may exist (depending on the optical lattice), 
though the name Kondo-Heisenberg model 
usually refers to the model with $V=0$.  
The model \eqref{eqn:SUN-KHM} will be discussed in Sec.~\ref{sec:SUSY} in the context of the boson-fermion 
supersymmetry. 
\subsection{Symmetries}
\label{sec:symmetry}
\subsubsection{U(1) and SU(N)}
\label{sec:U1-SUN}
Now we discuss several symmetries of the models \eqref{eqn:SUN-KLM-2} and \eqref{eqn:SUN-KHM} 
which will be important in the following discussion.  
First of all, they are invariant under the following 
(site-independent) U(1) gauge transformation for the itinerant fermion:
\begin{equation}
c_{i}^{\alpha} \to \be^{i \phi} c_{i}^{\alpha}   \; ,
\label{eqn:U1-gauge}
\end{equation}
which is associated with the conservation of the total fermion number: 
\[  
\mathcal{N}_{\text{c}} = \sum_{i} \sum_{\alpha=1}^{N} c_{\alpha,\,i}^\dag c^{\alpha}_{i} \; .
\]
On top of the above U(1) symmetry, the two models are invariant under the SU($N$) transformation:
\begin{equation}
\begin{split}
& c_{i}^{\alpha} \to \sum_{\beta=1}^{N} U(\boldsymbol{\theta})^{\dagger}_{\alpha\beta} c_{i}^{\beta}      \\
& S^{A}_{i}  \to \sum_{B=1}^{N^{2}-1} S_{i}^{B}  [R_{\text{adj}} (\boldsymbol{\theta}) ]_{BA}  \; ,
\end{split}
\end{equation}
where the transformation $U(\boldsymbol{\theta}) \in \text{SU($N$)}$ is defined by
\[
U(\boldsymbol{\theta}) := \exp \left( -i \sum_{A=1}^{N^{2}-1} \theta_{A} G^{A} \right)  \; ,
\]
and the adjoint representation $R_{\text{adj}}(\boldsymbol{\theta}) \in \text{SO($N^{2}-1$)}$ is related to 
$U$ as:
\[
\begin{split}
& R_{\text{adj}} (\boldsymbol{\theta}) := \exp \left( -i \sum_{A=1}^{N^{2}-1}  \theta_{A} G_{\text{adj}}^{A} \right)     \\
& \left( [G_{\text{adj}}^{A} ]_{BC} := -i f^{ABC} , \;  [G_{\text{adj}}^{A} ]^{\text{T}} = - G_{\text{adj}}^{A}  \right)   \; .
\end{split}
 \] 
 The invariance of the Hamiltonians \eqref{eqn:SUN-KLM-2} and \eqref{eqn:SUN-KHM} 
 can be seen if we note that the $\widehat{\mathcal{U}}$ transforms the fermion ``spin'' as:
\begin{equation}
\hat{s}_{i}^{A}  
\to  \sum_{B=1}^{N^{2}-1} \hat{s}_{i}^{B}  [R_{\text{adj}} (\boldsymbol{\theta}) ]_{BA} \; .
\end{equation}
As in SU(2), the SU($N$)-symmetry leads to several conserved quantities.  
First, in SU($N$), there are $(N-1)$ commuting generators that play the role of $S^{z}$, and correspondingly,  
we have a set of the $(N-1)$ conserved quantities called weight; each state in a given SU($N$) multiplet 
has a unique weight [the converse is not true for $N \geq 3$; see Ref.~\cite{Georgi-book-99} for more details 
on SU($N$)].  
In this paper, we denote the local weight associated with $\hat{s}_{i}^{A} + S_{i}^{A}$ and its sum over 
the entire system by $\lambda_{i}$ and $\Lambda_{\text{tot}}$, respectively.   
Giving the local weight $\lambda_{i}$ and the local fermion number $n_{i}+1$ 
($1$ from the localized $e$ fermion, $0 \leq n_{i} \leq N$) is equivalent to specifying the color-resolved fermion density 
$n_{\alpha,i} =\sum_{m=g,e} c_{m\alpha,\,i}^{\dagger}c_{m\alpha,\,i}$ ($\alpha=1,\ldots, N$).  
On top of the weight, there are $N-1$ Casimir operators $\mathcal{C}_{2},\ldots, \mathcal{C}_{N}$ 
that are the SU($N$) analog of the spin squared $\mathbf{S}^{2}$.  
Among them, the quadratic Casimir $\mathcal{C}_{2}$ defined in Appendix ~\ref{sec:Kondo-energy} is crucial in 
evaluating the Kondo energies.  

\subsubsection{Particle-hole transformation}
\label{sec:PH-tr}
\begin{figure}[htb]
\begin{center}
\includegraphics[scale=0.7]{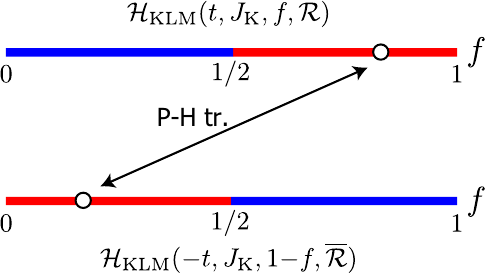}
\end{center}
\caption{Particle-hole transformation for SU($N$) Kondo lattice model \eqref{eqn:SUN-KLM-2}.  
When the local moments are not self-conjugate, SU($N$) Kondo lattice model with local moments $\mathcal{R}$ 
is mapped onto {\em another}  model in which the local moments are replaced 
with the conjugate ones $\overline{\mathcal{R}}$. 
\label{fig:P-H-tr}}
\end{figure}

The particle-hole (P-H) transformation that interchanges the creation and annihilation operators for the itinerant fermions: 
\begin{equation}
c^{\alpha}_{i} \leftrightarrow c^{\dagger}_{\alpha,i}  \quad (\alpha=1,\ldots, N) \; 
\label{eqn:P-H-tr}
\end{equation}
plays a key role to understand the global phase structure of the $N=2$ KLM.  
The first hallmark of the SU($N \geq 3$) KLM is the {\em absence} of the particle-hole symmetry, as we will see below. 
By the P-H transformation \eqref{eqn:P-H-tr}, physical quantities transform as: 
\begin{subequations}
\begin{align}
& n_{i} = \sum_{\alpha} n_{i,\uparrow}  \xrightarrow{\text{P-H}}    N - n_{i} \, \quad 
\left( f := n_{\text{c}}/N  \xrightarrow{\text{P-H}}  1-f   \right)  
 \label{eqn:P-H-tr-1}  \\
& \hat{s}_{i}^{A}   \xrightarrow{\text{P-H}}     
\sum_{\sigma,\sigma^{\prime}} c_{\beta,i}^{\dagger} [- (G^{A})^{\text{T}}]_{\beta\alpha} c^{\alpha}_{i}  
=: \hat{\bar{s}}_{i}^{A}
\label{eqn:P-H-tr-2} \\
& \left( c_{\alpha,i}^{\dagger} c^{\alpha}_{j} + c_{\alpha,j}^{\dagger} c^{\alpha}_{i} \right)  
\xrightarrow{\text{P-H}}   
 -  \left( c_{\alpha,i}^{\dagger} c^{\alpha}_{j} + c_{\alpha,j}^{\dagger} c^{\alpha}_{i} \right)    \; .
 \label{eqn:P-H-tr-3}
\end{align}
\end{subequations}
The third equation implies that the P-H transformation flips the sign of the hopping term 
\begin{equation}
t_{i,j} \xrightarrow{\text{P-H tr}}  - t_{i,j} \; ,
\label{eqn:P-H-tr-4}
\end{equation}
which is not important on bipartite lattices as we can always undo the minus sign 
by applying the gauge transformation \eqref{eqn:U1-gauge} with $\phi=\pi$ only on one of the sublattices.  
The equation \eqref{eqn:P-H-tr-2} tells that the SU($N$)-spin (of the itinerant fermions) $\hat{\mathbf{s}}_{i}$  
maps onto its 
conjugate: 
\[ \hat{\mathbf{s}}_{i} \xrightarrow{\text{P-H tr}} \hat{\bar{\mathbf{s}}}_{i} \; ,  \] 
which in general is different from $\hat{\mathbf{s}}_{i}$ for $N \geq 3$ (see Appendix \ref{sec:Young-diag} for 
the conjugate representations).  
Therefore, the Kondo coupling changes its form by the particle-hole transformation:
\begin{equation}
\sum_{A=1}^{N^{2}-1}\hat{s}_{i}^{A} S_{i}^{A} 
\xrightarrow{\text{P-H tr}} 
\sum_{A=1}^{N^{2}-1} \hat{\bar{s}}_{i}^{A} S_{i}^{A}  \; \left( \neq   \sum_{A=1}^{N^{2}-1} \hat{s}_{i}^{A} S_{i}^{A}  \right)  
\end{equation}
for $N \geq 3$ \footnote{%
When $N=2$, we can apply a unitary transformation $i \sigma^{y}$ to $(c_{\uparrow},c_{\downarrow})^{\text{T}}$ 
to make $\hat{\bar{s}} \to \hat{s}$.}.   
However, if we simultaneously replace the local spin with its conjugate
\begin{equation}
S_{i}^{A} \xrightarrow{\text{conjugate}}  \overline{S}_{i}^{A}  =  - (S_{i}^{A})^{\text{T}}   \; ,
\end{equation}
the Kondo coupling changes to
\begin{equation}
\sum_{A=1}^{N^{2}-1}\hat{s}_{i}^{A} S_{i}^{A} \xrightarrow{\text{P-H}}  \sum_{A=1}^{N^{2}-1}\hat{\bar{s}}_{i}^{A} S_{i}^{A}  
\xrightarrow{\text{conjugate}} \sum_{A=1}^{N^{2}-1} \hat{\bar{s}}_{i}^{A} \overline{S}_{i}^{A}  \; .
\end{equation}
As the quadratic Casimirs for an irreducible representation (``spin'') $\mathcal{R}$ and its 
conjugate $\overline{\mathcal{R}}$ are the same,  
the two different Kondo couplings 
$\sum_{A} \hat{s}_{i}^{A} S_{i}^{A}$ and $\sum_{A} \hat{\bar{s}}_{i}^{A} \overline{S}_{i}^{A}$ share the same set of the eigenvalues.  
Therefore, the particle-hole transformation relates the SU($N$) KLM  at filling $f$ with the local spins $\mathcal{R}$  
to the same model at filling $1-f$ with the conjugate local spins $\overline{\mathcal{R}}$ (see Fig.~\ref{fig:P-H-tr}):
\begin{equation}
\mathcal{H}_{\text{KLM}} (t_{i,j},J_{\text{K}} , f; \mathcal{R}) \; 
\overset{\text{P-H tr.}}{\Longleftrightarrow} \; 
\mathcal{H}_{\text{KLM}} ( - t_{i,j} ,J_{\text{K}} , 1-f ; \overline{\mathcal{R}}) 
\end{equation}
which means that the ground state of the KLM with local moments in $\mathcal{R}$ at filling $f$ is obtained from 
that of {\em another} KLM with local moments $\overline{\mathcal{R}}$ at filling $1-f$ by particle-hole transformation, 
and vice versa.  
Only for self-conjugate local spins ($\overline{\mathcal{R}}=\mathcal{R}$), particle-hole symmetry exists guaranteeing  
the symmetry of the phase diagram with respect to the half-filling $f=1/2$.  
\section{Strong-coupling limits}
\label{sec:strong-coupling}
In this section, we derive the effective Hamiltonians describing the low-energy physics 
of the SU($N$) KLM \eqref{eqn:SUN-KLM-2} in the limit of large $| J_{\text{K}} |$.  
Throughout this and the next section \ref{sec:ferromagnetism}, we only consider the pure KLM  
\eqref{eqn:SUN-KLM-2} [or the model \eqref{eqn:SUN-KHM} with $J_{\text{H}}=V=0$].   
\subsection{Strong-coupling ground state}
\label{sec:strong-coupling-GS}
\begin{figure}[htb]
\begin{center}
\includegraphics[width=\columnwidth,clip]{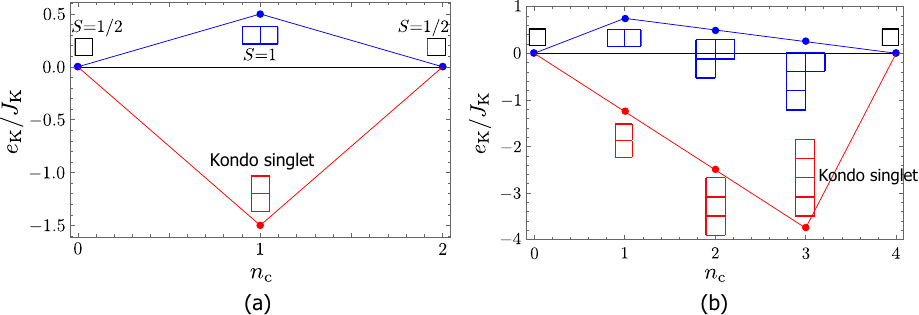}
\end{center}
\caption{Kondo energies \eqref{eqn:Kondo-energies-SUN} vs. fermion number $n_{\text{c}}$ ($0 \leq n_{\text{c}} \leq N$) 
for (a) $N=2$ and (b) $N=4$ 
(red for anti-symmetric $(n_{\text{c}}+1)$-tensor representation).  Note that when $N=2$ [i.e., SU(2)], the Kondo energy is 
symmetric with respect to $n_{\text{c}}=N/2=1$ as a consequence of the particle-hole symmetry, while this symmetry is lost 
for $N \geq 3$.  
\label{fig:Kondo-energies}}
\end{figure}

To carry out the strong-coupling ($t/J_{\text{K}}$) expansion, it is necessary to first evaluate the local Kondo energy 
\begin{equation}
J_{\text{K}} \sum_{A=1}^{N^{2}-1}\hat{s}_{i}^{A} S_{i}^{A} =:  J_{\text{K}} \, \hat{\mathbf{s}}_{i}{\cdot} \mathbf{S}_{i} \; .
\label{eqn:def-Kondo-energy}
\end{equation}
As in the standard SU(2) case, the value depends on the fermion number $n = \sum_{\alpha} c_{\alpha}^{\dagger} c_{\alpha}$ 
and how the fermion spin $\hat{\mathbf{s}}_{i}$ and the local moment $\mathbf{S}_{i}$ 
are combined into the total SU($N$) spin.   
When the fermion number is $n_{\text{c}}$ ($n_{\text{c}}=1,\ldots,N-1$), the following two ``total spins'' are possible 
[see Eq.~\eqref{eqn:CG-decomp-SUN-KLM}]: 
\begin{equation}
\text{\scriptsize $n_{\text{c}}{+}1$} \left\{ 
{\tiny \yng(1,1,1,1)  }
\right. 
\quad , \quad 
\text{\scriptsize $n_{\text{c}}$} \left\{ 
{\tiny \yng(2,1,1)  }
\right.   
\quad (1 \leq n_{\text{c}} \leq N-1)   \; .
\end{equation}
When $n_{\text{c}} =0$ (empty) and $n_{\text{c}}=N$ (fully-occupied), the itinerant fermions are in the SU($N$)-singlets  
and only the local spin contributes to the total spin:
\begin{equation}
\underset{\text{fermions}}{ \bullet }
\otimes 
\underset{\text{local moment}}{\yng(1)}
\;  \sim  \; \;
 \yng(1) 
\quad (\text{$\boldsymbol{N}$-rep.})   \; .
\end{equation}
The Kondo energies $e_{\text{K}} (n_{\text{c}})$ for these states are calculated using the quadratic Casimir $\mathcal{C}_{2}$ introduced in 
Appendix~\ref{sec:Kondo-energy}: 
\begin{equation}
\begin{split}
& e_{\text{K}} (n_{\text{c}}) = -\frac{N+1}{N} n_{\text{c}} J_{\text{K}} \quad \text{for } \;\;  
{\tiny \text{\scriptsize $n_{\text{c}}{+}1$} \left\{ 
\yng(1,1,1,1) \right.   } 
\; , \\
& e_{\text{K}} (n_{\text{c}}) = \left( 1-\frac{n_{\text{c}}}{N} \right) J_{\text{K}} \quad \text{for } \;\;  
{\tiny   \text{\scriptsize $n_{\text{c}}$} \left\{ 
\yng(2,1,1)
\right. }
\quad (1 \leq n_{\text{c}} \leq N-1)    \\
& e_{\text{K}} (n_{\text{c}}) = 0 \quad (n_{\text{c}}=0,N) 
\; .
\end{split}
\label{eqn:Kondo-energies-SUN}
\end{equation}    
The results are summarized in Table \ref{tab:local-states}.  
The Kondo energies are plotted in Fig.~\ref{fig:Kondo-energies} for $N=2$ and $N=4$ 
against the fermion number (per site) $n_{\text{c}}$.  
When $N=2$, the energy is symmetric with respect to $n_{\text{c}}=1$ (half-filling) reflecting 
the particle-hole symmetry, while for $N = 4$, this symmetry is lost.   
It is well-known \cite{Tsunetsugu-S-U-97} that, for SU(2), a single fermion (electron) and an $S=1/2$ moment 
can form a spin-singlet called the Kondo singlet.  
However, a local spin in the $\boldsymbol{N}$ (${\tiny \yng(1)}$) representation cannot be screened by a single fermion, 
and in fact we need $N-1$ fermions to make an SU($N$) singlet [see Fig.~\ref{fig:Kondo-energies}(b)].    
With this caution in mind, we take over the name {\em Kondo singlet} 
to denote this SU($N$) singlet state formed by $(N-1)$ fermions and a local spin.    

Now let us determine the strong-coupling ground state by minimizing the total Kondo energy 
$\sum_{i} J_{\text{K}} \, \hat{\mathbf{s}}_{i}{\cdot} \mathbf{S}_{i}$.  
To this end, we start from the reference state in which the local fermion number $n_{\text{c}}$ is integer and uniform.  
For $1 \leq n_{\text{c}} \leq N-1$, the states in ${\tiny \text{\scriptsize $n_{\text{c}}{+}1$} \left\{ 
\yng(1,1,1,1) \right.   }$  and  ${\tiny   \text{\scriptsize $n_{\text{c}}$} \left\{ \yng(2,1,1)\right. }$ are selected {\em at each site}  
for $J_{\text{K}} >0$ and $J_{\text{K}} <0$, respectively.     
Naively, the strong-coupling ground state may be obtained by uniformly tiling one of these two according to the sign of $J_{\text{K}}$.  
However, this strategy works only when $n_{\text{c}} = N-1$ (for $J_{\text{K}}>0$) or $n_{\text{c}} =1$ (for $J_{\text{K}} <0$) 
at which the Kondo energy $e_{\text{K}} (n_{\text{c}})$ is concave (see Fig.~\ref{fig:Kondo-energies}).    
The linear behavior of the Kondo energy for other $n_{\text{c}}$ means that 
we can move one fermion from one site to another [$(n_{\text{c}}, n_{\text{c}}) \to (n_{\text{c}} +1, n_{\text{c}}-1 )$] 
without changing the Kondo energy of the entire system.  Repeating this procedure, we can generate many {\em inhomogeneous} 
ground states which are degenerate with the uniform one.  Physically, we may expect that the (strong-coupling) ground states 
at these commensurate fillings $f \,(=n_{\text{c}}/N)$ are metallic.  

In what follows, we restrict ourselves to the two regions 
\begin{equation}
\begin{split}
& \text{(i)} \; 0 \leq f \leq 1/N  \quad (J_{\text{K}} < 0) \\
& \text{(i)} \; 1-1/N \leq f \leq 1 \quad (J_{\text{K}} > 0) 
\end{split}
\end{equation}
in which the strong-coupling ground state is well under control.   
Specifically, in the case (i), each site is occupied either by ${\tiny \yng(1)}$ ($n_{\text{c}}=0$)  
or by ${\tiny \yng(2)}$ ($n_{\text{c}}=1$) 
as is shown in Fig.~\ref{fig:SUN-large-Jk-GS-FM}(a,b), while in (ii), only the Kondo singlet $\bullet$ ($n_{\text{c}}=N-1$) and  
the $N$-dimensional ``spin'' ${\tiny \yng(1)}$ ($n_{\text{c}}=N$) appear in the ground states 
[see Fig.~\ref{fig:SUN-large-Jk-GS-AF}(a,c)].    
In particular, when $J_{\text{K}}<0$, the (spin-degenerate) ground states at the commensurate filling $f=1/N$ ($n_{\text{c}}=1$ 
fermion at each site) 
are given by a uniform tiling of the state ${\tiny \yng(2)}$ [Fig.~\ref{fig:SUN-large-Jk-GS-FM}(a)].   
When $J_{\text{K}} >0$, on the other hand, 
the ground state at $f=1-1/N$ ($n_{\text{c}}=N-1$) is the tensor product of the Kondo singlets 
[Fig.~\ref{fig:SUN-large-Jk-GS-AF}(a)] and is non-degenerate.   
\begin{table}[htb]
\caption{\label{tab:local-states} List of all the $2^{N} \times N$ on-site states and their Kondo energies.}
\begin{ruledtabular}
\begin{tabular}{lccc}
fermion num. & total spin & Kondo energy  & degeneracy  \\
\hline
$n_{\text{c}}=0$ (empty) & ${\tiny \yng(1)}$ & $0$ & $N$ \\
\hline
$1 \leq n_{\text{c}} \leq N{-}1$
& ${\tiny \text{\scriptsize $n_{\text{c}}$} \left\{  \yng(2,1,1) \right. }$ &  $\left( 1-\frac{n_{\text{c}}}{N} \right) J_{\text{K}} $   
& $ \frac{n_{\text{c}} (N{+}1)!}{(n_{\text{c}} {+} 1)!(N{-} n_{\text{c}})!} $        \\
  & ${\tiny \text{\scriptsize $n_{\text{c}}{+}1$} \left\{  \yng(1,1,1,1) \right.  }$  &  $-\frac{N+1}{N} n_{\text{c}} J_{\text{K}} $ 
& $ \frac{N!}{(n_{\text{c}} {+}1)!(N{-} n_{\text{c}} {-}1)!} $       \\
\hline
$n_{\text{c}}=N$ (fully-occupied) & ${\tiny \yng(1)}$ & $0$ & $N$ \\
\end{tabular}
\end{ruledtabular}
\end{table}

\subsection{Ferromagnetic Kondo coupling}
\label{sec:FM-Kondo}
\subsubsection{Insulating phase at $f=1/N$ ($n_{\text{c}}=1$)}
\begin{figure}[htb]
\begin{center}
\includegraphics[width=\columnwidth,clip]{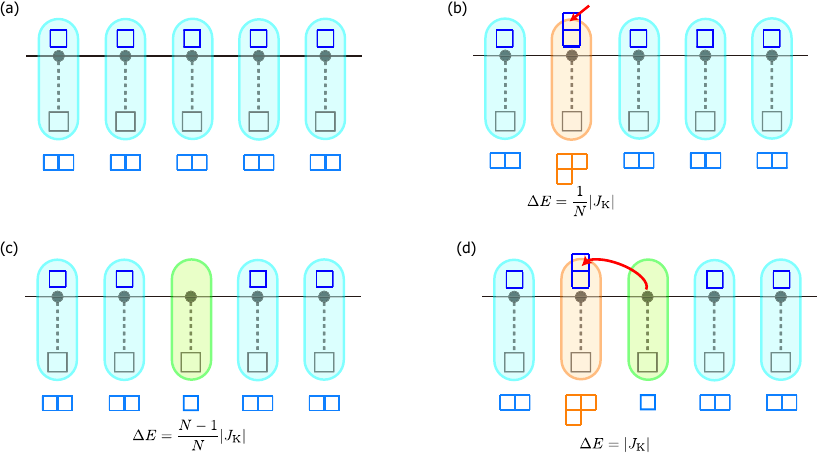}
\end{center}
\caption{Ground and excited states in the strong-coupling limit when the Kondo interaction is ferromagnetic ($J_{\text{K}} <0$). 
(a) Singlet ground state, (b) single-particle excitation, (c) ``hole'' excitation, and (d) particle-hole excitation.   
Red ovals denote ``Kondo singlets''. 
\label{fig:SUN-large-Jk-GS-FM}
}
\end{figure}

Having found the ground states at $t=0$, let us consider the excited states at the commensurate filling 
$f=1/N$.  
The energy cost by the addition of one fermion [Fig.~\ref{fig:SUN-large-Jk-GS-FM}(b)] is calculated as
\begin{equation}
\Delta_{\text{c}}^{+} = - J_{\text{K}}/N = |J_{\text{K}}| /N \; ,
\end{equation}
while, when a fermion is removed [see Fig.~\ref{fig:SUN-large-Jk-GS-FM}(c)], the cost is given by: 
\begin{equation}
\Delta_{\text{c}}^{-} = - (N-1)J_{\text{K}}/N = (N-1)|J_{\text{K}}| /N \; .
\end{equation}   
The two gaps are different for $N \geq 3$ as a consequence of the absence of particle-hole symmetry.  
To move one fermion from one site to another [Fig.~\ref{fig:SUN-large-Jk-GS-FM}(d)], we need extra energy:
\begin{equation}
\Delta_{\text{c}}^{\text{P-H}} = 
(1-2/N)J_{\text{K}} - 2 \times (1-1/N)J_{\text{K}} = - J_{\text{K}} = | J_{\text{K}}|  \; .
\end{equation}
The three energies satisfy:
\begin{equation}
\Delta_{\text{c}}^{+} \leq \Delta_{\text{c}}^{-} < \Delta_{\text{c}}^{\text{P-H}} 
\end{equation}
(the equalities hold when $N=2$).   
Therefore, we may expect that an insulating ground state forms at $1/N$-filling (i.e., $n_{\text{c}}=1$).  

One may naively create the spin excitation by turning one of the ${\tiny \yng(2)}$ spins  
in Fig.~\ref{fig:SUN-large-Jk-GS-FM}(a) into ${\tiny \yng(1,1)}$.  However, this may not be the lowest spin excitation.  
The behavior of the SU($N$) spin-sector is non-trivial as the strong-coupling ($t=0$) ground state is highly 
degenerate (the degree of degeneracy is $[N(N+1)/2]^{L}$) with respect to the SU($N$) spin states.  
The second-order degenerate perturbation in $t$ yields the following effective SU($N$) Heisenberg Hamiltonian 
for the spin sector:
\begin{equation}
\mathcal{H}_{\text{eff}} = \frac{t^{2}}{2 |J_{\text{K}}|} 
\sum_{i} \mathcal{S}^{A}_{i} ({\tiny \yng(2)}) \mathcal{S}^{A}_{i+1} ({\tiny \yng(2)})   \; ,
\label{eqn:Heisenberg-in-sym-rank-2}
\end{equation}
with the spins $\mathcal{S}_{i}$ belonging to the symmetric rank-2 tensor ${\tiny \yng(2)}$ 
[when $N=2$, \eqref{eqn:Heisenberg-in-sym-rank-2} reduces to the spin-1 Heisenberg chain].   
The inclusion of the $J_{\text{H}}$-interaction merely renormalizes the coupling: 
$\frac{t^{2}}{2 |J_{\text{K}}|} \to \frac{t^{2}}{2 |J_{\text{K}}|} + J_{\text{H}}/4$.  
According to the recent analytical and numerical studies \cite{Lecheminant-15,Wamer-L-M-A-20,Nataf-G-M-21}, 
the low-energy physics of the model \eqref{eqn:Heisenberg-in-sym-rank-2} depends on the parity of $N$.  
That is, the strong coupling SU($N$) Kondo lattice model \eqref{eqn:SUN-KLM-2} at filling $f=1/N$  
is a spin-gapped insulator when $N$ is even, while it is an insulator with algebraic spin correlations when $N$ is odd.  
Except when $N=2$, these spin-gapped insulators for $N=\text{even}$ are not the symmetry-protected topological phases  
associated with PSU($N$) \cite{Duivenvoorden-Q-13}.  
\subsubsection{Effective Hamiltonian for $0 \leq f \leq 1/N$}
\label{sec:eff-Ham-FM}
\begin{figure}[htb]
\begin{center}
\includegraphics[width=\columnwidth,clip]{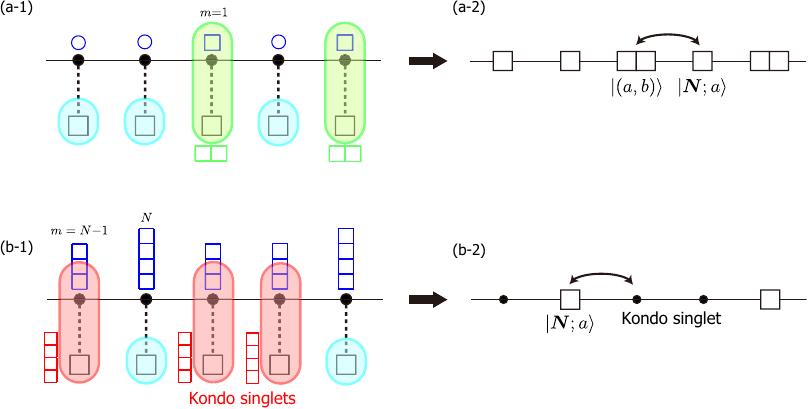}
\end{center}
\caption{Ground states in the strong-coupling limit at partial filling $0 \leq n_{\text{c}} \leq 1$ when $J_{\text{K}}<0$ (a-1) 
and $N-1 \leq n_{\text{c}} \leq N$ when $J_{\text{K}}>0$ (b-1).  The ground-state subspace is spanned 
by $\boldsymbol{N}$ and the rank-2 symmetric tensor (Kondo singlets $\bullet$ and $\boldsymbol{N}$) in (a) [(b)].  
The corresponding effective Hamiltonians within the ground-state subspaces are shown in (a-2) and (b-2). 
\label{fig:SUN-large-Jk-GS-incom}}
\end{figure}

When we move away from the commensurate filling $f=1/N$, the strong-coupling 
ground states now contain a certain number of sites in ${\tiny \yng(1)}$ ($n_{\text{c}}=0$) 
as well as those in ${\tiny \yng(2)}$ [see Fig.~\ref{fig:SUN-large-Jk-GS-incom}(a-1,2)].  
These ground states are highly degenerate with respect not only to 
the locations of the ${\tiny \yng(1)}$ spins in the ``sea'' of ${\tiny \yng(2)}$ but also 
to the SU($N$) spin states at the individual sites.  
This huge degeneracy might be partially or fully resolved by the motion of the fermions.  

To understand how the degeneracy is lifted, let us derive an effective Hamiltonian within the ground-state subspace 
by the first-order perturbation in $t$.  
To this end, we need to find the expression of the hopping term projected onto the ground-state manifold.  
To begin with, we explicitly write the expressions of the spin states in ${\tiny \yng(1)}$ and ${\tiny \yng(2)}$.  
As the fermionic (F) part of the states with $n_{\text{c}}=0$ (empty) and $n_{\text{c}}=1$ are given respectively by: 
\begin{equation}
\begin{split}
& n_{\text{c}}=0: \quad 
 |0\rangle_{\text{F}}  \\
& n_{\text{c}}=1: \quad 
|{}_{\alpha}\rangle_{\text{F},i}  := c_{\alpha,i}^{\dagger}  |0\rangle_{\text{F}}  \quad (\alpha=1,\ldots,N) \; , 
\end{split}
\end{equation}
the two types of spin states are given by (``S'' stands for the local-spin part of the state):
\begin{subequations}
\begin{equation}
|\boldsymbol{N}; \alpha \rangle_{i} = |{}_{\alpha} \rangle_{\text{S},i} \otimes |0\rangle_{\text{F}} 
\quad (n_{\text{c}}=0, \text{${\tiny \yng(1)}$-spin})
\label{eqn:basis-N-rep-FM-Jk}
\end{equation}
and 
\begin{equation}
\begin{split}
& | (\alpha, \beta) \rangle_{i}  \\
& = 
\begin{cases}
 |{}_{\alpha} \rangle_{\text{S},i} \otimes c^{\dagger}_{\alpha,i}|0\rangle_{\text{F}}  & (\alpha=\beta) \\
 \frac{1}{\sqrt{2}} \left\{  |{}_{\alpha} \rangle_{\text{S},i} \otimes c^{\dagger}_{\beta,i}  |0\rangle_{\text{F},i} 
 + |{}_{\beta} \rangle_{\text{S},i} \otimes c^{\dagger}_{\alpha,i}  |0\rangle_{\text{F},i}  \right\}  
 & (a < \beta) 
 \end{cases}  \\
 & \quad (n_{\text{c}}=1,   \text{${\tiny \yng(2)}$-spin} )   \; .
 \end{split} 
 \label{eqn:basis-rank-2-symmetric}
 \end{equation}
 \end{subequations}
 
When no fermion occupies a pair of adjacent sites $(i,i+1)$, the hopping term simply annihilates the state. 
If each of the pair is occupied by one fermion (i.e.,  ${\tiny \yng(2) }-{\tiny \yng(2) }$), the action of the hopping always 
creates excited states:
\begin{equation}
\begin{split}
&  |\, {\tiny \yng(2) }\, \rangle_{i} \otimes | \, {\tiny \yng(2) } \, \rangle_{i+1} 
 \xrightarrow{c_{\mu,i+1}^{\dagger} c_{i}^{\mu}} 
|\, {\tiny \yng(1)} \, \rangle_{i} \otimes \left| \, {\tiny \yng(2,1) } \, \right\rangle_{i+1}   
\quad  (\Delta E = | J_{\text{K}} | )   \\ 
&  |\, {\tiny \yng(2) }\, \rangle_{i} \otimes | \, {\tiny \yng(2) } \, \rangle_{i+1} 
 \xrightarrow{c_{\mu,i}^{\dagger} c_{i+1}^{\mu}} 
\left| \, {\tiny \yng(2,1) } \, \right\rangle_{i}  \otimes  |\, {\tiny \yng(1)} \, \rangle_{i+1}  
\quad  (\Delta E = | J_{\text{K}} | )   
\end{split}
\end{equation}
and does not contribute to the first-order perturbation.   Therefore, only the pairs of the form ${\tiny \yng(2) }-{\tiny \yng(1)}$ 
or ${\tiny \yng(1) }-{\tiny \yng(2)}$ can contribute to the effective Hamiltonian.  
To investigate the action of the hopping operator onto the above pair, it is convenient to use 
the following expressions of the fermion operators $\tilde{c}_{i}^{\alpha}$ and $\tilde{c}_{\alpha,i}^{\dagger}$ 
projected onto the ground-state manifold: 
 \begin{equation}
 \begin{split}
 & \tilde{c}_{i}^{\alpha} = |\boldsymbol{N}; \alpha \rangle_{i} \langle (\alpha,\alpha)|_{i}  \\
& \phantom{ \tilde{c}_{i}^{\alpha}= } 
+ \frac{1}{\sqrt{2}} \sum_{\beta < \alpha} |\boldsymbol{N}; \beta\rangle_{i} \langle(\beta,\alpha)|_{i} 
 + \frac{1}{\sqrt{2}} \sum_{\beta > \alpha} |\boldsymbol{N}; \beta\rangle_{i} \langle(\alpha,\beta)|_{i}   \\
 & \tilde{c}_{\alpha,i}^{\dagger} = | (\alpha,\alpha)\rangle_{i} \langle \boldsymbol{N}; \alpha |_{i}  \\
 & \phantom{  \tilde{c}_{\alpha,i}^{\dagger} = } 
 + \frac{1}{\sqrt{2}} \sum_{\beta < \alpha} | (\beta,\alpha)\rangle_{i} \langle \boldsymbol{N}; \beta |_{i} 
 + \frac{1}{\sqrt{2}} \sum_{\beta > \alpha} | (\alpha,\beta)\rangle_{i} \langle \boldsymbol{N}; \beta |_{i}   \;  .
 \end{split}
 \label{eqn:FM-Jk-projected-fermion}
 \end{equation}  
 \begin{widetext}
\noindent%
In writing the effective Hamiltonian, we also need to keep track of the sign factors arising from 
 the fermion exchange.  Fortunately, for open boundary conditions, no extra sign appears in the subspace considered here, 
 and we obtain the following effective Hamiltonian:
 \begin{equation}
 \begin{split}
& - t \sum_{\mu=1}^{N} \tilde{c}^{\dagger}_{\mu, i} \tilde{c}^{\mu}_{j} 
\underbrace{ |\boldsymbol{N}; \alpha \rangle_{i} }_{{\tiny \yng(1)}} \otimes 
 \underbrace{ | (\beta,\gamma) \rangle_{j} }_{{\tiny \yng(2)}} \\
 &= 
 \begin{cases}
 \displaystyle{%
 - \frac{t}{\sqrt{2}} |(\alpha, \beta)\rangle_{i} \otimes |\boldsymbol{N}; \beta \rangle_{j} 
-  \frac{t}{\sqrt{2}} ( \sqrt{2} - 1 ) \delta_{\alpha\beta} |(\alpha,\alpha)\rangle_{i} \otimes |\boldsymbol{N}; \alpha \rangle_{j}  
 }    & \text{when $\beta = \gamma$} \\
\begin{split}
& - \frac{t}{2}  |(\alpha,\gamma)\rangle_{i} \otimes |\boldsymbol{N}; \beta\rangle_{j} 
- \frac{t}{2} |(\alpha, \beta)\rangle_{i} \otimes |\boldsymbol{N}; \gamma \rangle_{j}  
  \\
& \quad - \frac{t}{2} (\sqrt{2}-1) \delta_{\alpha \gamma} |(\alpha,\alpha)\rangle_{i} \otimes |\boldsymbol{N}; \beta \rangle_{j} 
- \frac{t}{2} (\sqrt{2}-1) \delta_{\alpha\beta} |(\alpha,\alpha)\rangle_{i} \otimes |\boldsymbol{N}; \gamma \rangle_{j} 
\end{split}
& \text{when $\beta \neq \gamma$}   \; .
\end{cases}
\end{split}
 \label{eqn:SUN-projected-hopping-FM-1D}
 \end{equation}
 Obviously, all the non-zero off-diagonal matrix elements appearing here are negative when $t > 0$.   
In the cold-atom realization, $t$ is positive and this condition is satisfied from the outset.     
 When the system is periodic, hopping across the boundary yields 
 a fermion sign $(-1)^{\mathcal{N}_{\text{c}}-1}$ that necessitates an additional condition $\mathcal{N}_{\text{c}}=\text{odd}$ 
for the non-positivity.   
 \end{widetext}
 
\subsection{Antiferromagnetic Kondo coupling}
\label{sec:AF-Kondo}
\subsubsection{Insulating phase at $f=1-1/N$}
\begin{figure}[htb]
\begin{center}
\includegraphics[width=\columnwidth,clip]{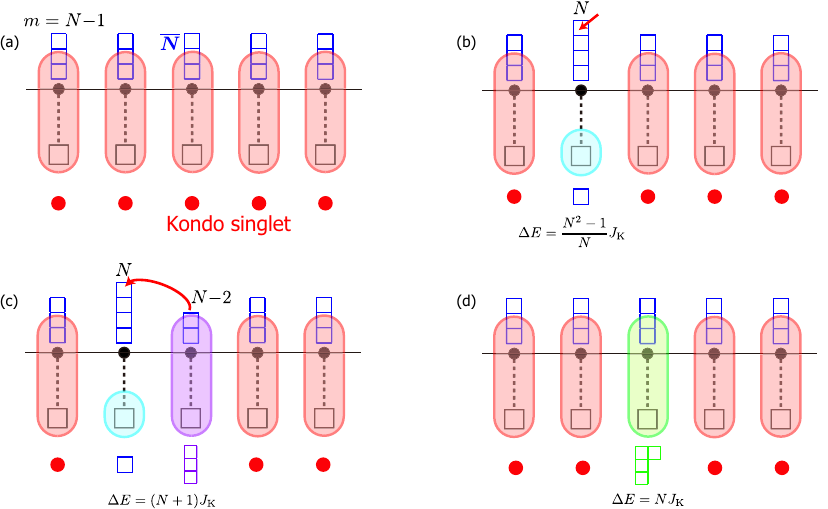}
\end{center}
\caption{Ground and gapped excited states in the strong-coupling limit ($J_{\text{K}} >0$). 
(a) Singlet ground state, (b) single-particle excitation, (c) particle-hole excitation, and (d) ``spin'' excitation.   
Red ovals denote ``Kondo singlets''. 
\label{fig:SUN-large-Jk-GS-AF}}
\end{figure}

As has been discussed in Sec.~\ref{sec:strong-coupling-GS}, 
the strong-coupling ground state when $f=1-1/N$ (or $n_{\text{c}}=N-1$ fermions 
at each site) and $J_{\text{K}}>0$ is the product of the local Kondo singlets shown  
in Fig.~\ref{fig:SUN-large-Jk-GS-AF}(a) and is non-degenerate.  
Adding (removing) one fermion to (from) this ground state costs finite energy [Fig.~\ref{fig:SUN-large-Jk-GS-AF}(b)]
\begin{equation}
\Delta_{\text{c}}^{+} = (N^{2}-1)J_{\text{K}}/N \quad  
[\Delta_{\text{c}}^{-} = (N+1)J_{\text{K}}/N]  \; .
\end{equation}   
The addition always costs more energy, i.e., $\Delta_{\text{c}}^{+} - \Delta_{\text{c}}^{-} = (N+1)(N-2)J_{\text{K}}/N >0$ when $N \geq 3$ (due to the absence of the particle-hole symmetry).   
On the other hand, moving one fermion from one site to another [particle-hole excitations; 
see Fig.~\ref{fig:SUN-large-Jk-GS-AF}(c)] also costs an energy 
\begin{equation}
\Delta_{\text{c}}^{\text{P-H}} = (N+1)J_{\text{K}}  \; .
\end{equation}
To create a ``spin'' excitation in the ground state, one needs to excite one of the SU($N$) Kondo singlets to the adjoint 
[see Fig.~\ref{fig:SUN-large-Jk-GS-AF}(d)]: 
\begin{equation}
{\tiny 
\text{\scriptsize $N$} \left\{  \yng(1,1,1,1) \right.  }  \to 
{\tiny 
\text{\scriptsize $N{-}1$} \left\{  \yng(2,1,1) \right.  }   \; .
\end{equation}
The energy cost of this ``magnetic'' excitation is $\Delta_{\text{s}} = N J_{\text{K}}$.   
These imply that for large enough $J_{\text{K}} \, (>0)$, the ground state of $\mathcal{H}_{\text{KLM}}$ \eqref{eqn:SUN-KLM-2} 
at $f=1-1/N$ is a spin-gapped insulator, which is the SU($N$) analog of the well-known Kondo insulator 
at half-filling in the SU(2) KLM \cite{Tsvelik-94}.  

\subsubsection{Effective Hamiltonian for $1-1/N \leq f \leq 1$}
\label{sec:eff-Ham-AF}
Now let us find the effective Hamiltonian which is first-order in $t$.  
The ground-state manifold in this region is spanned by the following two types of fermion states:
\begin{equation}
\begin{split}
& n_{\text{c}}=N: \quad 
 |\text{f}\rangle_{\text{F},i} := \prod_{\beta=1}^{N} c^{\dagger}_{i,\beta} |0\rangle_{\text{F}}  
 = c^{\dagger}_{i,1} \cdots c^{\dagger}_{i,N} |0\rangle_{\text{F}} \\
& n_{\text{c}}=N-1: \quad 
|{}^{\alpha}\rangle_{\text{F},i} := c_{i}^{\alpha} |\text{f}\rangle_{\text{F},i} 
=  (-1)^{\alpha-1} \prod_{\beta \neq \alpha} c^{\dagger}_{i,\beta} |0\rangle_{\text{F}}     \; . 
\end{split}
\label{eqn:2-low-energy-states-high-density}
\end{equation}
Due to the strong $J_{\text{K}}>0$, the above states with $n_{\text{c}}=N-1$ (conjugate $\overline{\boldsymbol{N}}$) 
and $n_{\text{c}} =N$ (singlet), 
together with the local moment in $\boldsymbol{N}$, form the SU($N$) multiplets in $\bullet$ (Kondo singlet) 
and $\boldsymbol{N}$ (${\tiny \yng(1)}$), respectively [see Fig.~\ref{fig:SUN-large-Jk-GS-incom}(b-2)].  
Written explicitly, these multiplets are given as:
\begin{subequations}
\begin{equation}
 |\bullet \rangle_{i} := \frac{1}{\sqrt{N}} \sum_{\alpha=1}^{N} |{}_{\alpha} 
 \rangle_{\text{S},i} \otimes |{}^{\alpha} \rangle_{\text{F},i} 
= \frac{1}{\sqrt{N}} \sum_{\alpha=1}^{N} |{}_{\alpha} \rangle_{\text{S},i} \otimes c_{i}^{\alpha} |\text{f}\rangle_{\text{F},i} 
\label{eqn:basis-Kondo-singlet}
\end{equation}
when $m=N-1$, and   
\begin{equation}
 |\boldsymbol{N}; \alpha \rangle_{i} := |{}_{\alpha} \rangle_{\text{S},i} \otimes  |\text{f}\rangle_{\text{F},i}  
 \quad (\alpha =1,\ldots, N)   \quad \text{when $m=N$}\; ,
\label{eqn:basis-N-rep-AF-Jk} 
\end{equation}
\end{subequations}
where $|{}_{\alpha} \rangle_{\text{S},i}$ denotes the states of the local spin in $\boldsymbol{N}$.  
When $n_{\text{c}}=N-1$, there is yet another multiplet in the adjoint representation 
$\left( {\tiny \text{\scriptsize $N{-}1$} \left\{  \yng(2,1,1) \right.  } \right)$:
\begin{equation}
|\text{adj};A\rangle_{i} 
= \sum_{\alpha,\beta=1}^{N} [ G^{A} ]^{\alpha}{}_{\beta} |{}_{\alpha} \rangle_{\text{S},i} 
\otimes |{}^{\beta} \rangle_{\text{F},i}   \; ,
\end{equation}
which is not allowed energetically in the ground state when $J_{\text{K}}$ is positively large 
but is necessary for considering the higher-order corrections in Sec.~\ref{eqn:3-site-processes}.   

It is straightforward to write the expressions of the fermion operators projected onto the subspace spanned by 
the above two states \eqref{eqn:basis-Kondo-singlet} and \eqref{eqn:basis-N-rep-AF-Jk}, which are given 
(up to the many-body fermion sign) by: 
\begin{equation}
\widetilde{c}^{\dagger}_{\alpha,i} = \frac{1}{\sqrt{N}} |\boldsymbol{N}; \alpha \rangle_{i} \langle\bullet |_{i}  \, , \quad 
\widetilde{c}_{i}^{\alpha} = \frac{1}{\sqrt{N}}  |\bullet \rangle_{i} \langle\boldsymbol{N}; \alpha |_{i}  \; .
\label{eqn:AF-Jk-projected-fermion}
\end{equation}
Note that they now obey the non-standard anti-commutation relations:
\begin{equation}
\begin{split}
& \{ \widetilde{c}_{i}^{\alpha} , \widetilde{c}_{i}^{\beta} \} 
= \{ \widetilde{c}^{\dagger}_{\alpha,i} , \widetilde{c}^{\dagger}_{\beta,i} \} = 0  \\
& \{ \widetilde{c}_{i}^{\alpha} , \widetilde{c}^{\dagger}_{\beta,i} \} = 
\frac{1}{N} \delta^{\beta}{}_{\alpha} |\bullet \rangle \langle\bullet |  
+ \frac{1}{N} |\boldsymbol{N}; \alpha \rangle \langle\boldsymbol{N}; \beta |  \; .
\end{split}
\label{eqn:AF-Jk-projected-ACR}
\end{equation}
We will see in Sec.~\ref{sec:SUSY} that these commutation relations are closely related to Bose-Fermi supersymmetry.  

If we use the following for the fermionic part of the many-body basis states
\begin{equation}
\begin{split}
& | i_{1}^{\alpha_{1}}, i_{2}^{\alpha_{2}}, \cdots, i_{n}^{\alpha_{n}} \rangle_{\text{F}}  \\
& = \prod_{k=1}^{n} (-1)^{(N-1)i_{k}} |\cdots\rangle_{\text{F}} 
\otimes |{}^{\alpha_{1}}\rangle_{\text{F},i_{1}} \otimes \cdots \otimes |{}^{\alpha_{2}}\rangle_{\text{F},i_{2}} \\
& \phantom{ = } \quad 
\otimes  \cdots \otimes |{}^{\alpha_{n}}\rangle_{\text{F},i_{n}} \otimes |\cdots \rangle_{\text{F}}  \; ,
\end{split}
\label{eqn:SUN-GS-basis-AF-1D}
\end{equation}
the effective Hamiltonian (up to the first order in $t$) is given by:
\begin{subequations}
\begin{align}
\begin{split}
& -t \sum_{\beta=1}^{N} \widetilde{c}^{\dagger}_{\beta,i+1} \widetilde{c}_{i}^{\beta} \, 
| \cdots\rangle \otimes | \boldsymbol{N}; \alpha\rangle_{i} \otimes | \bullet \rangle^{\prime}_{i+1} 
\otimes | \cdots \rangle \\
& = - \frac{t}{N} | \cdots\rangle \otimes | \bullet \rangle^{\prime}_{i} \otimes 
 | \boldsymbol{N}; \alpha \rangle_{i+1} \otimes | \cdots \rangle \; , 
 \end{split}
  \label{eqn:SUN-projected-hopping-1D-1}
 \\
 \begin{split}
& -t \sum_{\beta=1}^{N} \widetilde{c}^{\dagger}_{\beta,i} \widetilde{c}_{i+1}^{\beta} \, 
| \cdots\rangle \otimes | \bullet \rangle^{\prime}_{i} \otimes | \boldsymbol{N}; \alpha \rangle_{i+1} \otimes | \cdots \rangle  \\ 
& = - \frac{t}{N} | \cdots\rangle \otimes | \boldsymbol{N}; \alpha \rangle_{i} 
 \otimes | \bullet \rangle^{\prime}_{i+1} \otimes | \cdots \rangle  \; , 
 \end{split}
 \label{eqn:SUN-projected-hopping-1D-2}
 \end{align}
\end{subequations} 
with $| \bullet \rangle_{i}^{\prime}$ now being defined by:
\begin{equation}
| \bullet \rangle^{\prime}_{i} := (-1)^{(N-1)i } \frac{1}{\sqrt{N}} \sum_{\alpha=1}^{N} |{}_{\alpha} \rangle_{\text{S},i} 
\otimes c_{i}^{\alpha} 
|\text{f}\rangle_{\text{F},i}   \; .
\label{eqn:Kondo-singlet-modified}
\end{equation}
The extra sign factor $(-1)^{(N-1)i }$ in Eqs.~\eqref{eqn:SUN-GS-basis-AF-1D} and \eqref{eqn:Kondo-singlet-modified} 
has been introduced to eliminate the many-body fermion sign.  
\section{Ferromagnetism}
\label{sec:ferromagnetism}
\begin{figure}[htb]
\begin{center}
\includegraphics[scale=0.45]{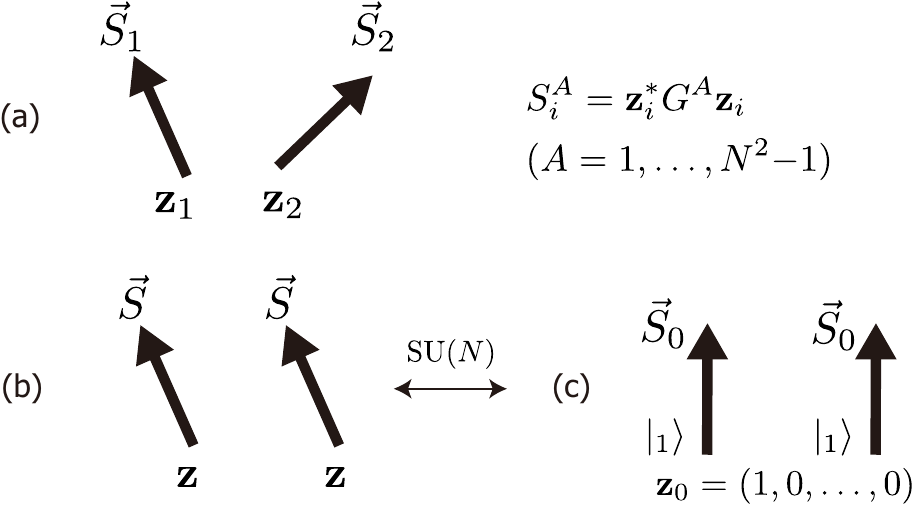}
\end{center}
\caption{Semiclassical (coherent-state) representation of the states $|\mathbf{z} \rangle$ of SU($N$) ``spins''  
(\raisebox{-0.2ex}{$\square$}) and the corresponding SU($N$) moment $\vec{S}$.  
Generic configuration (a) and ferromagnetic ones (b,c).    
The SU($N$) ``spin'' moment is represented by an $(N^{2}-1)$-dimensional (real) vector 
$\vec{S} = (S^{1} ,\ldots, S^{N^{2}-1} )$ given by  
$S^{A} = \mathbf{z}^{\dagger} G^{A} \mathbf{z}$ [see Eq.~\eqref{eqn:SUN-gen-by-z}].  
\label{fig:ferro-by-spinor}}
\end{figure}

In this section, we prove that the ground state of the SU($N$) KLM \eqref{eqn:SUN-KLM-2} is ferromagnetic 
in the two regions considered in the previous section.  
Before doing so, we first characterize the ferromagnetic states in SU($N$)-symmetric systems in physical terms.  
In the ordinary SU(2)-symmetric systems, the standard intuitive picture of ferromagnetic states is that all the spins 
depicted as arrows are aligned in a particular direction.  In general, this simple picture holds only in SU(2) where  
all the spin-$S$ can be represented by the (symmetrized) product of $2S$ spin-1/2s and 
any $S=1/2$ states can be uniquely represented by points on the unit sphere.  
In SU($N$), even a pair of SU($N$) ``spins'' that are coupled ferromagnetically may not point in the same direction.   
Nevertheless, in the situations considered here (i.e., $N$-component fermions coupled to 
$N$-component local spins), we can use, instead of the three-component unit vector in SU(2),  
a complex unit vector $\mathbf{z}=(z_{1},\ldots,z_{N})$ to uniquely specify the state of a single $N$-component fermion 
(for both itinerant and localized fermions)  \footnote{%
In fact, the state space of an $N$-component fermion is isomorphic to $\mathbb{C}P^{N-1}$.}; 
the coincidence of $\mathbf{z}$ up to phases means the same SU($N$) spin state and  the same spin 
$\langle S^{A} \rangle$ [see Fig.~\ref{fig:ferro-by-spinor}(b) and Eq.~\eqref{eqn:SUN-gen-by-z}].  
As will be seen in Sec.~\ref{sec:SUN-double-exchange}, the coherent state $|\mathbf{z}\rangle$ 
specified by the complex vector $\mathbf{z}$ 
coincides with the ``fully-polarized'' state $|{}_{1}\rangle = c_{1}^{\dagger}|0\rangle$ up to SU($N$) rotation.  
Therefore, we may think of the ferromagnetic state in the present cases as the one 
in which {\em all} the constituent SU($N$) spins ${\tiny \yng(1)}$ (both itinerant and local; 
the spins ${\tiny \yng(2)}$ are treated as made of two ${\tiny \yng(1)}$'s) 
are in the same state, e.g., $|{}_{1}\rangle$ [$\mathbf{z}=(1,0,\ldots,0)$; see Fig.~\ref{fig:ferro-by-spinor}(c)]  
and those obtained from it by applying the SU($N$) lowering operators.  
\subsection{Ferromagnetism in low-density region $0 < f < 1/N$}
\label{sec:ferro-FM-Kondo}
As we have seen, the strong-coupling effective Hamiltonian \eqref{eqn:SUN-projected-hopping-FM-1D} for 
large ferromagnetic $J_{\text{K}}$ is given by a non-positive matrix when $t > 0$.   
Of course, as the effective Hamiltonian \eqref{eqn:SUN-projected-hopping-FM-1D} preserves 
the total SU($N$) weight, the full effective Hamiltonian 
for an $L$-site system decomposes into several blocks with respect to the conserved weight $\Lambda_{\text{tot}}$.  
In the following, we consider one of those blocks with a given total SU($N$) weight $\Lambda_{\text{tot}}$, 
which we denote by $\mathcal{H}_{\text{eff}}(\Lambda_{\text{tot}}; L)$.  
To prove ferromagnetism, we need one more important property called irreducibility or indecomposability 
on top of the non-positivity.   A given real Hamiltonian matrix $\mathcal{H}$ is said irreducible if 
there exists a sequence of non-zero off-diagonal matrix elements 
$\mathcal{H}_{i k_{n-1}} \mathcal{H}_{k_{n-1} k_{n-2}} \cdots \mathcal{H}_{k_{1}j}$ ($n \geq 1$) 
for any pair of $(i,j)$ ($i \neq j$).   Physically, this implies 
that repeated application of $\mathcal{H}$ can connect any pair of the initial ($j$) and final ($i$) states.  
For simplicity of the argument, we assume the open boundary condition \footnote{%
In periodic systems, the matrix elements for the hopping {\em across} the boundary acquire an additional 
fermion sign $(-1)^{\mathcal{N}_{\text{c}}-1}$ ($\mathcal{N}_{\text{c}}$ is the total fermion number in the system).  
Due to this factor, the condition for positivity depends explicitly on the parity of the fermion number.}.  
Then, it is straightforward to show, by the mathematical induction, that the block Hamiltonian 
$\mathcal{H}_{\text{eff}}(\Lambda_{\text{tot}}; L)$ 
is irreducible (see Appendix~\ref{sec:indecomposability} for the sketch of the proof).  

Now, by the Perron-Frobenius theorem (see, e.g., Ref.~\cite{Tasaki-book-20} for a physicist-friendly exposition of the theorem), 
we can prove that the ground state $|\Psi_{\text{g.s.}}(\Lambda_{\text{tot}}) \rangle$ 
of $\mathcal{H}_{\text{eff}}(\Lambda_{\text{tot}}; L)$ is unique 
and is given by a superposition of all the possible tensor-products of the states \eqref{eqn:basis-N-rep-FM-Jk} 
and \eqref{eqn:basis-rank-2-symmetric} (allowed for the value of $\Lambda_{\text{tot}}$) with 
strictly positive coefficients.   Obviously, the ferromagnetic states [with the same SU($N$) weight $\Lambda_{\text{tot}}$] 
have a similar sign property, which implies 
\begin{equation}
\mathcal{P}_{\text{ferro}}(\Lambda_{\text{tot}})  |\Psi_{\text{g.s.}}(\Lambda_{\text{tot}}) \rangle \neq 0 
\end{equation}
with $\mathcal{P}_{\text{ferro}}(\Lambda_{\text{tot}})$ being the projector onto the ferromagnetic states 
in the subspace with $\Lambda_{\text{tot}}$.  
Since the block Hamiltonian $\mathcal{H}_{\text{eff}}(\Lambda_{\text{tot}}; L)$ commutes with $\mathcal{P}_{\text{ferro}}(\Lambda_{\text{tot}})$, 
we immediately see that 
$\mathcal{P}_{\text{ferro}}(\Lambda_{\text{tot}})  |\Psi_{\text{g.s.}}(\Lambda_{\text{tot}}) \rangle$ is a ground state 
of $\mathcal{H}_{\text{eff}}(\Lambda_{\text{tot}}; L)$, which is allowed, by the uniqueness, if and only if 
$|\Psi_{\text{g.s.}}(\Lambda_{\text{tot}}) \rangle \propto 
\mathcal{P}_{\text{ferro}}(\Lambda_{\text{tot}})  |\Psi_{\text{g.s.}}(\Lambda_{\text{tot}}) \rangle$, i.e., 
the unique ground state is ferromagnetic.  
This generalizes the rigorous statement for the SU(2) model in Ref.~\cite{Kubo-82} to arbitrary $N$.  
\subsection{Ferromagnetism in high-density region}
\label{sec:ferro-AF-Kondo}
\subsubsection{Perculiarity in 1D}
\label{sec:ferromagnetism-1D-AF}
Now let us consider the effective Hamiltonian for $J_{\text{K}} >0$ which describes the dynamics of 
the mobile ${\tiny \yng(1)}$ spins in the background of the Kondo singlets $\bullet$.  
The first-order effective Hamiltonian \eqref{eqn:SUN-projected-hopping-1D-1} and \eqref{eqn:SUN-projected-hopping-1D-2} 
in one dimension (1D) [which we denote by $\mathcal{H}^{(1)}$; see Fig.~\ref{fig:SUN-large-Jk-GS-incom}(b-2)] 
has non-positive off-diagonal matrix elements when $t>0$.  
However, when the open boundary condition is chosen, these off-diagonal elements 
simply exchange an adjacent pair of a ${\tiny \yng(1)}$-spin and a Kondo singlet $\bullet$ 
without changing the background SU($N$) spin configurations $\{ \alpha_{k} \}$ \footnote{%
When the periodic boundary condition is imposed, on the other hand, cyclic permutation of the spin configurations 
is allowed \cite{Caspers-I-89}.  However, it is clear that cyclic permutations do not connect all the allowed states 
which still prevents the application of the Perron-Frobenius theorem.  In fact, for finite periodic systems, 
states other than ferromagnetic ones can have lower energies.}, and the lowest-order effective Hamiltonian $\mathcal{H}^{(1)}$   
does not stabilize any particular magnetic orders, as is well-known in the one-dimensional Hubbard model 
at $U=\infty$ \cite{Caspers-I-89,Ogata-S-90}.   
Nevertheless, the effective Hamiltonian $\mathcal{H}^{(1)}$ partially resolves the positional degeneracy 
thereby reducing the degree of degeneracy  
${\tiny \begin{pmatrix} L \\ \mathcal{N}_{\square} \end{pmatrix}} \times N^{\mathcal{N}_{\square}}$ 
[with $\mathcal{N}_{\square}= \mathcal{N}_{\text{c}}-(N-1)L$ being the number of ${\tiny \yng(1)}$-spins]   
down to $N^{\mathcal{N}_{\square}}$;     
the full $\mathcal{H}^{(1)}$ decomposes into $N^{\mathcal{N}_{\square}}$ {\em identical} diagonal blocks 
each of which describes the hopping of $\mathcal{N}_{\square}$ non-interacting spinless fermions.   
By the Perron-Frobenius theorem, the ground state of each block (with a given fixed spin configuration 
$\{ \alpha_{k} \}$) is unique and constructed by summing up all the possible states with the same sequence 
$\{ \alpha_{k} \}$ over the positions of the $\mathcal{N}_{\square}$ fermions with strictly positive coefficients.  

This peculiar situation in one dimension is a natural consequence of the equivalence 
between the one-dimensional SU($N$) KLM \eqref{eqn:SUN-KLM-2} and the one-dimensional SU($N$) Hubbard model: 
\begin{equation}
\mathcal{H}_{\text{Hubbard}} = - t_{\text{H}} \sum_{i} 
\sum_{\alpha=1}^{N} \left\{ c^{\dagger}_{\alpha,i} c_{i+1}^{\alpha} + \text{H.c.} \right\} 
+ U \sum_{i} n_{i} (n_{i} -1)   \;  ,
\label{eqn:1-band-SUN-Hubbard}
\end{equation}   
which generalizes the known equivalence \cite{Lacroix-85} in SU(2) to SU($N$).  
Specifically, the $U=\infty$ effective Hamiltonian of the model \eqref{eqn:1-band-SUN-Hubbard} 
with $t_{\text{H}} = t/N$ for filling $0 \leq f \leq 1/N$ coincides with that of 
the $J_{\text{K}}=\infty$ SU($N$) KLM for filling $1-1/N \leq f \leq 1$ [\eqref{eqn:SUN-projected-hopping-1D-1} and 
\eqref{eqn:SUN-projected-hopping-1D-2}].   

The above equivalence still holds with the identification $t_{\text{H}} = t/N$ 
even in higher dimensions if we treat the ${\tiny \yng(1)}$-spins as fermions (see Sec.~\ref{sec:SUSY-AF-JK} for more 
details).  
However, the non-positivity is non-trivial since we now have complicated fermion sign factors in front of $t$.  
These sign factors are under control, e.g.,  
when there is only one hole (i.e., $\mathcal{N}_{\text{c}} =N \mathcal{N}_{\Lambda}-1$ with $\mathcal{N}_{\Lambda}$ 
being the number of lattice sites).   
In the Hubbard language, the fermion number $\mathcal{N}_{\text{c}}^{(\text{KLM})} = N \mathcal{N}_{\Lambda}-1$ 
corresponds to $\mathcal{N}_{\text{c}}^{(\text{Hubbard})} =\mathcal{N}_{\Lambda}-1$, i.e., one less fermion from $1/N$-filling,  
at which we expect the SU($N$) analog of 
the Nagaoka's ferromagnetism to occur for $t_{\text{H}} < 0$ \cite{Katsura-T-13,Bobrow-S-L-18} 
if the lattice structure is properly chosen.  
Therefore, we can borrow the results in the Hubbard model to show that the ground state of the SU($N$) Kondo lattice 
model in dimensions greater than 1 is ferromagnetic when there is exactly one hole 
$\mathcal{N}_{\text{c}} =N \mathcal{N}_{\Lambda}-1$ and the lattice satisfies certain conditions.  
\subsubsection{Higher-order corrections}
\label{eqn:3-site-processes}
The first-order effective Hamiltonian $\mathcal{H}^{(1)}$ only resolves the degeneracy 
in the positions of the ${\tiny \yng(1)}$-spins leaving the $N^{\mathcal{N}_{\square}}$-fold spin degeneracy intact.  
To lift the huge SU($N$)-spin degeneracy in the $J_{\text{K}}=\infty$ KLM,  
we need to go to higher orders in $t$.  
We follow the strategy of Ref.~\cite{Sigrist-T-U-R-92} and consider the second-order effective Hamiltonian 
within the (smaller) subspace consisting of the ground states of the first-order Hamiltonian $\mathcal{H}^{(1)}$.   

\begin{figure}[htb]
\begin{center}
\includegraphics[width=\columnwidth,clip]{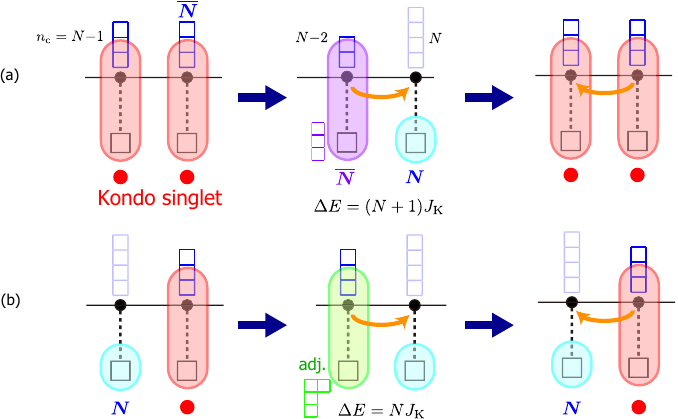}
\end{center}
\caption{Two types of second-order processes occurring on a two-site pair $\bullet{-}\bullet$ or $\boldsymbol{N}{-}\bullet$.  
For the $\boldsymbol{N}{-}\boldsymbol{N}$ pair, second-order processes are forbidden. 
The results can be compactly written as a two-body interaction \eqref{eqn:2-site-process-effective-int}.  
\label{fig:2-site-process}}
\end{figure}

We begin with the second-order processes involving two neighboring sites (see Fig.~\ref{fig:2-site-process}). 
It is easy to see that these $t^{2}$ corrections are all diagonal:
\begin{subequations}
\begin{align}
&  |\boldsymbol{N},\alpha \rangle_{i} \otimes  |\boldsymbol{N},b\rangle_{i+1} \longrightarrow 0 \\
& |\boldsymbol{N},\alpha \rangle_{i} \otimes |\bullet\rangle_{i+1} \longrightarrow 
-  \left( 1 - \frac{1}{N^{2}} \right) \frac{t^{2}}{N J_{\text{K}}} |\boldsymbol{N},\alpha \rangle_{i} \otimes |\bullet\rangle_{i+1}  \\
& |\bullet\rangle_{i} \otimes |\bullet\rangle_{i+1}  \longrightarrow 
- \frac{N-1}{N+1} \frac{2t^{2}}{N J_{\text{K}}} |\bullet\rangle_{i} \otimes |\bullet\rangle_{i+1} \; ,
\end{align}
\end{subequations}
from which we can read off the effective interactions:
\begin{equation}
\begin{split}
& \frac{2 (N-1) (2 N+1) t^2}{(N+1) N^3 J_{\text{K}}} \sum_{i} n_{i} ({\tiny \yng(1)}) n_{i+1} ({\tiny \yng(1)})  \\
& + \frac{2 (N-1) \left(N^2 - 2 N - 1\right) t^2}{(N+1)N^3 J_{\text{K}}} 
\sum_{i} \left\{ n_{i} ({\tiny \yng(1)}) - \frac{N^2}{N^2 - 2 N - 1} \right\} 
\end{split}
\label{eqn:2-site-process-effective-int}
\end{equation}
with $n_{i} ({\tiny \yng(1)})=1$ ($=0$) when the site $i$ is occupied by ${\tiny \yng(1)}$ ($\bullet$).   
Therefore, to find off-diagonal processes, we need to consider three-site processes.  

\begin{figure}[htb]
\begin{center}
\includegraphics[width=\columnwidth,clip]{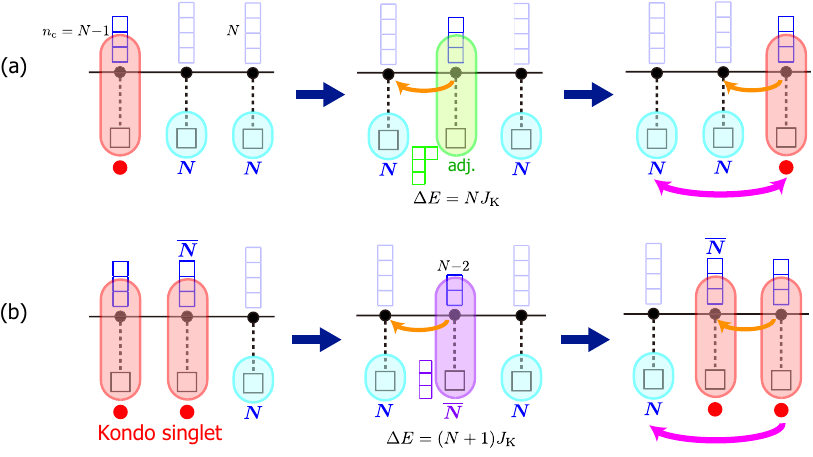}
\end{center}
\caption{Second-order processes occurring on three consecutive sites.   
In the first type, the Kondo singlet moves to the next-nearest-neighbor (NNN) site with the help of the fermion 
in the middle site.  This type of hopping includes (a) the 3-site processes that can change the spin states 
of the two $\boldsymbol{N}$s involved [see \eqref{eqn:process-1}] and 
(b) those moving the Kondo singlet without changing the background spin configurations [Eq.~\eqref{eqn:process-2}].  
The second type is just NNN (correlated) hopping of $\boldsymbol{N}$.   
\label{fig:3-site-process}}
\end{figure}

 At the order of $t^{2}$, only two types of three-site processes are possible (see Fig.~\ref{fig:3-site-process}): 
 \begin{subequations}
 \begin{align}
 \begin{split}
\text{(i)} \;  & \overset{N-1}{\bullet} - \overset{N}{\square} - \overset{N}{\square} \;\;  \overset{t}{\Rightarrow} \; \; 
 \overset{N}{\square} - \overset{N-1}{\tiny \yng(2,1,1,1)} (\text{adj.})- \overset{N}{\square} \\ 
 & \overset{t}{\Rightarrow} \;  \; 
 \overset{N}{\square} - \overset{N}{\square} - \overset{N-1}{\bullet} \quad (\Delta E = N J_{\text{K}})  \, , 
 \end{split}
 \\
\begin{split} 
\text{(ii)} \; & \overset{N-1}{\bullet} - \overset{N-1}{\bullet} - \overset{N}{\square}   
\;\; \overset{t}{\Rightarrow} \;\;  
\overset{N}{\square} - \overset{N-2}{\tiny \yng(1,1,1,1)} (\overline{\boldsymbol{N}}) -  \overset{N}{\square}  \\
& \overset{t}{\Rightarrow}  \;\;  
\overset{N}{\square} - \overset{N-1}{\bullet}- \overset{N-1}{\bullet} \quad  [ \Delta E = (N+1) J_{\text{K}} ]
 \; .
 \end{split}
\end{align}
\end{subequations}
The corresponding matrix elements read as:
\begin{subequations}
\begin{align} 
\begin{split}
& \text{(i)} \;\; 
| \bullet, N_{\alpha},N_{\beta} \rangle  \\
& \phantom{  \text{(i)} \;\;  }
\longrightarrow 
- \frac{t^{2}}{N^{2}J_{\text{K}}} |N_{\beta}, N_{\alpha}, \bullet \rangle  
+ \frac{t^{2}}{N^{3}J_{\text{K}}} |N_{\alpha}, N_{\beta}, \bullet \rangle 
\end{split}
\label{eqn:process-1}
 \\
& \text{(ii)} \;\; 
| \bullet, \bullet, N_{\alpha} \rangle \longrightarrow 
\frac{N-1}{N^{2} (N+1) J_{\text{K}} } t^{2} |N_{\alpha} , \bullet, \bullet \rangle \; ,
\label{eqn:process-2}
\end{align}
\end{subequations}
where $N_{\alpha}$ is the short-hand notation for $| \boldsymbol{N}; \alpha\rangle$.  
In the first term in \eqref{eqn:process-1}, a Kondo singlet $\bullet$ and $N_{\beta}$ interchange their 
positions with the help of $N_{\alpha}$ sitting in the middle [Fig.~\ref{fig:3-site-process}(a)], 
while in the second, a Kondo singlet 
$\bullet$ just goes through $N_{\alpha}$ and $N_{\beta}$ 
without disturbing the spin configurations.  Obviously, the type-(ii) processes \eqref{eqn:process-2}  
do not change the spin configurations [Fig.~\ref{fig:3-site-process}(b)]. 

Now we are at the place of constructing the (spin-only) effective Hamiltonian for the $N^{\mathcal{N}_{\square}}$-dimensional 
subspace of the spin-degenerate ground states.  
Let us evaluate the matrix elements of the first term in \eqref{eqn:process-1} that changes the spin configurations.  
Using the properties of the ground states (of $\mathcal{H}_{1}$) mentioned in Sec.~\ref{sec:ferromagnetism-1D-AF}, 
we readily see that the off-diagonal matrix elements of the second-order Hamiltonian are all non-positive, 
which allows us to apply the Perron-Frobenius theorem again to prove that the ground state of the Kondo lattice model 
\eqref{eqn:SUN-KLM-2} for sufficiently large $J_{\text{K}} (>0)$ is ferromagnetic when $t > 0$.    
\begin{figure}[htb]
\begin{center}
\includegraphics[scale=0.8]{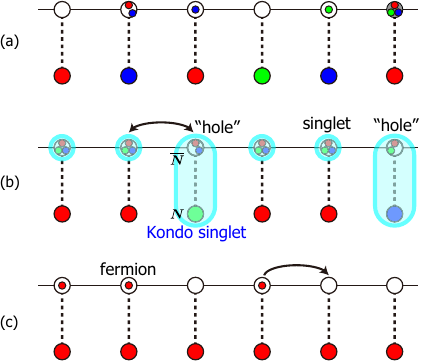}
\end{center}
\caption{Ferromagnetic ground states established here (for $N=3$).  
(a) A generic configuration and (b) the ferromagnetic ground state 
for $J_{\text{K}} >0$.  For the total ``hole'' number $0 \leq \mathcal{N}_{\text{h}} \leq L$ ($\mathcal{N}_{\text{h}}
= NL-\mathcal{N}_{\text{c}}$), 
the same number of the local spins are quenched by forming the Kondo singlets and only the surviving 
$L - \mathcal{N}_{\text{h}}$ unquenched moments (colored in red) form the ferromagnetic state.   
(c) Ferromagnetic ground state for $J_{\text{K}} <0$ and total fermion number $\mathcal{N}_{\text{c}}$ 
($0 \leq \mathcal{N}_{\text{c}} \leq L$),    
in which {\em both} $\mathcal{N}_{\text{c}}$ itinerant and $L$ local spins participate in ferromagnetism. 
\label{fig:SUN-ferro-config}}
\end{figure}
\subsection{Effects of residual interactions}
\label{sec:effect-residual-int}
A few remarks are in order about the effects of several terms 
that existed in the original two-orbital model \eqref{eqn:2-orbital-Hubbard} but are not explicitly taken into account 
in the KLM Hamiltonian \eqref{eqn:SUN-KLM-2}.   
First, we note that what is crucial to the proof of ferromagnetism is the non-positivity of the off-diagonal matrix elements.  
Therefore, the conclusion does not change even if we add any kind of diagonal terms (e.g., a non-uniform on-site potential 
associated with the harmonic trap) as far as they do not conflict with the prerequisites of the strong-coupling expansion. 

Also, in deriving the SU($N$) KLM \eqref{eqn:SUN-KLM-2} in Sec.~\ref{sec:two-orbital-Hubbard}, 
we have tentatively dropped the Hubbard $U^{(g)}$ interaction among the itinerant $g$-fermions 
\[   U^{(g)} n_{i} (n_{i} - 1)/2    \]
which may modify the effective Hamiltonians derived in Secs.~\ref{sec:FM-Kondo} and \ref{sec:AF-Kondo}.  
When $-J_{\text{K}} \,(>0)$ is large enough and $0 \leq f \leq 1/N$, we may keep only the $n_{i}=0,1$ states in which 
the Hubbard interaction is identically zero.  
When $J_{\text{K}}$ is antiferromagnetically large and $1-1/N \leq f \leq 1$, only the states with 
$n_{i} =N-1$ and $N$ are retained.  Even in this case, we see that the Hubbard-$U^{(g)}$ is totally irrelevant if we note:
\begin{equation}
\begin{split}
& \frac{1}{2} U^{(g)} n_{i} (n_{i} - 1)   \\
& = \frac{1}{2} U^{(g)} \{ n_{i} -(N-1)\} (n_{i} - N) + (N-1) U^{(g)} n_{i} \\
& \phantom{ = } - \frac{1}{2} (N-1) N U^{(g)}  \; .
\end{split}
\end{equation}
Therefore, we may conclude that none of the residual terms in \eqref{eqn:2-orbital-Hubbard} 
destabilizes the SU($N$) ferromagnetic phases found above.  
Of course, when these terms are comparable to or larger than 
the Kondo coupling $J_{\text{K}}$ which is assumed to be the dominant energy scale 
here, we expect other phases, e.g., conventional density-wave phases and more exotic symmetry-protected topological 
phases, to be stabilized \cite{Bois-C-L-M-T-15,Capponi-L-T-16}.   
\subsection{SU($N$) double exchange}
\label{sec:SUN-double-exchange}
In the previous sections, we have rigorously shown that the strong-coupling ground state of the SU($N$) 
KLM is ferromagnetic in certain regions of the phase diagram.  
To understand its mechanism simply, 
we try to generalize the double-exchange mechanism of ferromagnetism put forward 
in Refs.~\cite{Zener-51,Anderson-H-55,deGennes-60} to SU($N$) fermion systems.  
In order to treat both the itinerant fermions and the local spins within the same framework, 
we regain the orbital indices $m=g,e$ ($g$ for the itinerant fermions and $e$ for the local spins) 
used in Sec.~\ref{sec:two-orbital-Hubbard}.   
We start from the ``fully-polarized'' state of a local moment (i.e., an immobile $e$ fermion):
\begin{equation}
|\psi_{0}\rangle_{\text{S},i} = c_{e1,i}^{\dagger} |0\rangle_{\text{S}} = |{}_{1}\rangle_{\text{S},i}
\end{equation}
which is the SU($N$) analog of $|\!\! \uparrow \rangle$ in SU(2).  
In fact, the above reference state is invariant under the $\text{U(1)}\times \text{U($N-1$)}$ (stabilizer) subgroup 
of U($N$) of the form:
\[  
\left( 
\begin{array}{c|ccc}
\be^{i\theta} & \mathbf{0}  \\
\hline
\mathbf{0} & U(N-1)
\end{array}
\right) \; .
\]
Therefore, only a subset of U($N$) that are parametrized by an $(N-1)$-dimensional complex row vector $\mathbf{q}$ as:
\begin{equation}
\begin{split}
& \widehat{\mathcal{U}} (\mathbf{q}) = \exp[ - i \widehat{M} (\mathbf{q}) ] \, ,  \\
& \widehat{M} (\mathbf{q}) = (c_{e1}^{\dagger},\ldots, c_{eN}^{\dagger} ) 
\left( 
\begin{array}{c|c}
0 & \mathbf{q} \\
\hline
\mathbf{q}^{\text{T}} & \mathbf{0}_{N-1} 
\end{array}
\right)  
\begin{pmatrix} c_{e1} \\ \vdots \\ c_{eN} \end{pmatrix}
\end{split}
\end{equation} 
changes the reference state.  
Roughly, the $(N-1)$-dimensional vector $\mathbf{q}$ plays the same role 
as the unit vector $\boldsymbol{\Omega}$ (or the azimuthal and polar angles) in the Bloch coherent state.  
The fermion operators in the ``rotated'' frame are:
\begin{equation}
\begin{split}
& \bar{c}_{m \alpha}^{\dagger} (\mathbf{q})
= \widehat{\mathcal{U}}(\mathbf{q}) c_{m\alpha}^{\dagger} \widehat{\mathcal{U}}^{\dagger}(\mathbf{q}) 
= \sum_{\beta=1}^{N} c_{m \beta}^{\dagger} [ U(\mathbf{q})]_{\beta\alpha}    \\
& \bar{c}_{m\alpha} (\mathbf{q})
= \widehat{\mathcal{U}}(\mathbf{q}) c_{m\alpha} \widehat{\mathcal{U}}^{\dagger}(\mathbf{q}) 
= \sum_{\beta=1}^{N} [ U^{\dagger} (\mathbf{q})]_{\alpha\beta}  c_{m\beta} \;\; (m=g,e)
\end{split}
\label{eqn:rotated-fermion}
\end{equation}
with the $N\times N$ unitary $U(\mathbf{q})$ given by
\[
U(\mathbf{q}) = \exp \left[ - i  
\left( 
\begin{array}{c|c}
0 & \mathbf{q} \\
\hline
\mathbf{q}^{\dagger} & \mathbf{0}_{N-1} 
\end{array}
\right) 
\right] \; .
\] 
Using these rotated operators, the coherent state of the local spin at site-$i$ is defined by
\begin{equation}
\begin{split}
|\mathbf{q}_{i} \rangle_{\text{S},i} & :=  \bar{c}_{e1,i}^{\dagger}  |0\rangle_{\text{S}}  
 =  \widehat{\mathcal{U}}(\mathbf{q}_{i}) |\psi_{0}\rangle_{\text{S},i} \\
 & = \sum_{\beta=1}^{N}  [ U(\mathbf{q}_{i})]_{\beta 1} |{}_{\beta} \rangle_{\text{S},i} 
 =: \sum_{\beta=1}^{N}  [ \mathbf{z}(\mathbf{q}_{i})]_{\beta} |{}_{\beta} \rangle_{\text{S},i} 
 \; .
 \end{split}
 \label{eqn:imp-coherent-state}
\end{equation}
The $N$-dimensional complex vector $\mathbf{z}(\mathbf{q}_{i})$, which appeared at the beginning of 
Sec.~\ref{sec:ferromagnetism}, is given by the first column vector of $U(\mathbf{q}_{i})$ 
and satisfies $| \mathbf{z}(\mathbf{q}_{i}) |=1$.   
This is the SU($N$)-generalization of the Bloch coherent state
\[
|\mathbf{\Omega}_{i} \rangle
=  \be^{-i \frac{\chi_{i}}{2} } \left\{ 
\be^{-i \frac{\phi_{i}}{2} } \cos\frac{\theta_{i}}{2}  |\! \uparrow\rangle 
+ \be^{+i \frac{\phi_{i}}{2} }  \sin\frac{\theta_{i}}{2}  | \! \downarrow \rangle  
\right\}
\]
in which the role of the reference state $|\psi_{0}\rangle$ is played by $|\! \uparrow\rangle$.  
Instead of the vector spin $\mathbf{\Omega}_{i}(\phi_{i},\theta_{i})$, 
we use the complex unit vector $\mathbf{z}(\mathbf{q}_{i})$ to specify  
the state $|\mathbf{q}_{i} \rangle_{\text{S},i}$ of the local SU($N$) moment at site-$i$.    
In the semi-classical approximation, we treat the complex vector $\mathbf{z}(\mathbf{q}_{i})$ as the classical variable 
which we can fix at will like the classical vector spin.  Specifically, we 
approximate the local (quantum) SU($N$) spins $S^{A}_{i}$ by a set of $c$-numbers 
\begin{equation}
S^{A}(\mathbf{q}_{i}) = \langle \mathbf{q}_{i} | S_{i}^{A} |\mathbf{q}_{i} \rangle 
= \mathbf{z}^{\dagger} (\mathbf{q}_{i}) G^{A} \mathbf{z}(\mathbf{q}_{i})
\label{eqn:SUN-gen-by-z}
\end{equation}
[with $G^{A}$ being the $N \times N$ SU($N$) generators defined in Sec.~\ref{sec:model}].    
The exchange interaction between two such semi-classical spins 
\[ 
\sum_{A} S^{A}(\mathbf{p}_{i})S^{A}(\mathbf{q}_{i}) 
= | \mathbf{z}^{\ast} (\mathbf{p}_{i}) {\cdot} \mathbf{z}(\mathbf{q}_{i})|^{2} - 1/N \; .
\]
attains its (exact) maximal value $(1-1/N)$ when $\mathbf{z}(\mathbf{p}_{i}) =\mathbf{z}(\mathbf{q}_{i})$ up to a phase, 
i.e., when the two spins are coupled ferromagnetically: $S^{A}(\mathbf{p}_{i})=S^{A}(\mathbf{q}_{i})$.  
Therefore, when $J_{\text{K}} <0$, the strong Kondo coupling forces the itinerant and local spins 
on the same site to be parallel to each other. 
For the antiferromagnetic $J_{\text{K}} \,(>0)$, on the other hand, the Kondo energy is minimized for 
{\em any} configurations satisfying $\mathbf{z}^{\ast} (\text{itinerant}) {\cdot} \mathbf{z}(\text{local})=0$.  
This implies that when $N \geq 3$, the direction of the itinerant spin is not determined even if we fix the local moment. 

Now let us consider the hopping term. 
To this end, working with the itinerant ($g$) fermion in the same $\mathbf{q}$-frame as the local spin is convenient.  
Clearly, the $\alpha=1$ component $\bar{c}_{g1,i}$ of the rotated fermions corresponds to the direction 
of the local spin $S^{A}(\mathbf{q}_{i})$.  
With the help of Eq.~\eqref{eqn:rotated-fermion}, we can express the original hopping term by the rotated fermions 
$\bar{c}_{g\alpha,i}(\mathbf{q}_{i})$, and the resulting expression contains all the possible hopping processes 
(including the off-diagonal ones) $\bar{c}_{g\alpha,i}^{\dagger} \bar{c}_{g\beta,i+1}$ with the matrix elements 
\[
-t \left[ U^{\dagger}(\mathbf{q}_{i}) U(\mathbf{q}_{i+1}) \right]_{\alpha\beta} \; .
\]
The effect of the strong ferromagnetic Kondo coupling is taken into account by 
keeping only the $\alpha=\beta=1$ component which is ``parallel'' to the local spin \footnote{%
When $J_{\text{K}}$ is antiferromagnetic, the condition of the minimal Kondo coupling 
$\mathbf{z}^{\ast}_{\text{c},i} {\cdot} \mathbf{z}(\mathbf{q}_{i})=0$ does not favor a particular direction 
of the fermion state $\mathbf{z}^{\ast}_{\text{c},i}$ and the following argument fails.}:
\begin{equation}
-t \sum_{i} \left\{  \left[ \mathbf{z}^{\dagger}(\mathbf{q}_{i}) {\cdot} \mathbf{z}(\mathbf{q}_{i+1}) \right]  
\bar{c}_{g1,i}^{\dagger} \bar{c}_{g1,i+1} + \text{H.c.} \right\}  \;  .
\end{equation}
The matrix element can also be written as the overlap $\langle \mathbf{q}_{i} | \mathbf{q}_{i+1} \rangle_{\text{S}}$ 
between the local-spin states \eqref{eqn:imp-coherent-state} on the neighboring sites.  
Clearly, the hopping amplitude $ |\mathbf{z}^{\dagger}(\mathbf{q}_{i}) {\cdot} \mathbf{z}(\mathbf{q}_{i+1}) |$ 
of the fermions parallel to the local SU($N$) spin $S^{A}(\mathbf{q}_{i})$ 
is optimized when $\mathbf{z}(\mathbf{q}_{i})=\mathbf{z}(\mathbf{q}_{i+1})$ 
up to a phase, i.e., when the system is ferromagnetic: $S^{A}(\mathbf{q}_{i})=S^{A}(\mathbf{q}_{i+1})$ 
[see Fig.~\ref{fig:ferro-by-spinor}(b)].   
This is the generalization of the double-exchange mechanism of ferromagnetism 
in Refs.~\cite{Zener-51,Anderson-H-55,deGennes-60} to SU($N$).  
Our proof for $J_{\text{K}}<0$ tells that this simple argument can be made rigorous in one dimension 
without relying on the semi-classical approximation.  

Difference between $\text{SU}(N=2)$ and $\text{SU}(N \geq 3)$ becomes manifest 
in the antiferromagnetic case $J_{\text{K}}>0$.  
In fact, when $N=2$, we can easily generalize the above mechanism to the case of antiferromagnetic $J_{\text{K}}$ 
by keeping itinerant fermions {\em anti-parallel} to the local moments.   However, as strong $J_{\text{K}}$ alone 
no longer fixes the relative direction between the itinerant fermion and the local moment for $N \geq 3$, 
the simple double-exchange scenario breaks down when $J_{\text{K}}>0$. Nevertheless, ferromagnetism 
occurs in the large-$J_{\text{K}}(>0)$ SU($N$) KLM as we have shown rigorously in Sec.~\ref{sec:ferro-AF-Kondo}.   
\section{Supersymmetry}
\label{sec:SUSY}
In Sec.~\ref{sec:strong-coupling}, we have seen that the strong-coupling effective Hamiltonian contains 
two local degrees of freedom: two types of mobile spins ${\tiny \yng(1)}$ and ${\tiny \yng(2)}$ when $J_{\text{K}}<0$, or  
$\bullet$ and ${\tiny \yng(1)}$ when $J_{\text{K}}>0$ [see Figs.~\ref{fig:SUN-large-Jk-GS-incom}(a-2) and (b-2)].   
In the case of $J_{\text{K}}>0$, for instance, each site can take two different SU($N$) states $\bullet$ (which may be 
viewed as a hole) and ${\tiny \yng(1)}$ (a particle).  
As has been mentioned there, the creation and annihilation operators associated with these two (particle and hole) states 
no longer obey the standard anti-commutation relations for fermions [see Eq.~\eqref{eqn:AF-Jk-projected-ACR}].   
The situation is more involved when $J_{\text{K}} <0$ as is expected from Eq.~\eqref{eqn:FM-Jk-projected-fermion}.  
Nevertheless, we can regard these mobile spins as a kind of fermionic particles, and the boson-fermion 
supersymmetry SU($N | 1$) provides us with a convenient framework to handle these particles in a unifying way 
[a quick summary of the super Lie algebra SU($N|1$) is given in Appendix \ref{sec:SUSY-SU_N1}].    

\subsection{Supersymmetric SU($N$) $t$-$J$ model}
\label{sec:SUSY-SUN-t-J}
The super Lie algebra SU($N | 1$) consists of $N^{2}$ bosonic [see Eq.~\eqref{eqn:suN1-bosonic-gen}] 
and $2N$ fermionic [Eq.~\eqref{eqn:suN1-fermionic-gen}] generators satisfying particular algebraic relations 
\eqref{eqn:suM1-CCR-1}--\eqref{eqn:suM1-CCR-4}.  
The state vector of the simplest $(N+1)$-dimensional irreducible representation decomposes into two parts;  
the first $N$ components and the last one of the state vectors  
correspond to two different SU($N$) representations $\boldsymbol{N}$ (${\tiny \yng(1)}$) and the singlet ($\bullet$), respectively  
[see Eq.~\eqref{eqn:suN1-bosonic-gen}].  The $N^{2}$ bosonic generators are all block-diagonal 
with respect to these two sectors, 
while the $2N$ fermionic ones bring about the transitions between them. 

The simplest SU($N | 1$)-symmetric interaction can be found by considering the quadratic Casimir \eqref{eqn:quad-Casimir} 
which is a super-Lie-algebraic analog of $\mathbf{S}^{2}$ in SU(2):
\begin{equation*}
\mathcal{C}_{2} = \sum_{A=1}^{N^{2}-1} \mathcal{S}^{A} \mathcal{S}^{A} 
- \frac{1}{N(N-1)} \mathcal{K}^{2} 
- \sum_{\alpha=1}^{N}  \left( 
\mathcal{Q}_{\alpha} \widetilde{\mathcal{Q}}_{\alpha} 
- \widetilde{\mathcal{Q}}_{\alpha} \mathcal{Q}_{\alpha} 
\right)   
 \; 
\end{equation*}
with the bosonic generators $\mathcal{X}=\mathcal{S}^{A}, \mathcal{K}$ ($A=1,\ldots,N^{2}-1$) and the fermionic ones 
$\mathcal{X}= \mathcal{Q}_{\alpha}, \widetilde{\mathcal{Q}}_{\alpha}$ ($\alpha=1,\ldots,N$) 
given respectively by Eqs.~\eqref{eqn:suN1-bosonic-gen} and \eqref{eqn:suN1-fermionic-gen}.  
Then, it is straightforward to write the SU($N | 1$)-symmetric interaction as:
\begin{equation}
\begin{split}
& \mathcal{C}_{2} (\mathcal{X}_{i}+ \mathcal{X}_{j})  \\
& = \mathcal{C}_{2} (\mathcal{X}_{i}) + \mathcal{C}_{2} (\mathcal{X}_{j}) 
- 2 \sum_{\alpha=1}^{N}  \left( 
\mathcal{Q}_{\alpha,i} \widetilde{\mathcal{Q}}_{\alpha,j} 
- \widetilde{\mathcal{Q}}_{\alpha,i} \mathcal{Q}_{\alpha,j} 
\right)  \\
& \phantom{=} 
+ 2 \left\{ 
\sum_{A=1}^{N^{2}-1} \mathcal{S}^{A}_{i} \mathcal{S}^{A}_{j} 
- \frac{1}{N(N-1)} \mathcal{K}_{i} \mathcal{K}_{j} 
\right\}  
\end{split}
\label{eqn:gen-SUN1-Hamiltonian} 
\end{equation}
where the constant $\mathcal{C}_{2} (\mathcal{X})= \frac{N(N-2)}{N-1}$.   
Depending both on the irreducible representation we use and 
on how the local degrees of freedom realize in specific physical systems, 
the Hamiltonian \eqref{eqn:gen-SUN1-Hamiltonian} describes different physical situations.  
For instance, if we use the simplest $(N+1)$-dimensional representation and 
identify the first $N$ and the last one components as describing a fermionic particle carrying the SU($N$)-spin ${\tiny \yng(1)}$ 
and a hole, respectively, $\mathcal{Q}_{\alpha},$ ($\widetilde{\mathcal{Q}}_{\alpha}$) creates  (annihilates) 
the ${\tiny \yng(1)}$-particle; the term $\mathcal{Q}_{\alpha,i} \widetilde{\mathcal{Q}}_{\alpha,j} 
- \widetilde{\mathcal{Q}}_{\alpha,i} \mathcal{Q}_{\alpha,j} $ simply expresses the hopping of the ${\tiny \yng(1)}$-spin, 
and the model \eqref{eqn:gen-SUN1-Hamiltonian} describes interacting (fermionic) ${\tiny \yng(1)}$-particles 
hopping in the background of holes ($\bullet$).  
Typically, this situation occurs in the $U=\infty$ SU($N$) Hubbard model.  
In this simplest realization, the model \eqref{eqn:gen-SUN1-Hamiltonian} is known as 
the supersymmetric SU($N$) $t$-$J$ model \cite{Schlottmann-92,Kawakami-SUN-93,Schlottmann-93}  
which is an SU($N$)-generalization of the usual supersymmetric $t$-$J$ model for $N=2$ \cite{Wiegmann-88,Bares-B-90}.  
Thanks to the exact solution, low-energy physics is well understood and is known to be described by 
the $N$-component $\text{U(1)}\times\text{SU($N$)}_{1}$ Tomonaga-Luttinger liquid \cite{Kawakami-SUN-93}.  
\subsection{Antierromagnetic Kondo coupling}
\label{sec:SUSY-AF-JK}
To realize a supersymmetric model in the strong-coupling limit of the antiferromagnetic SU($N$) KLM, 
we first need to identify the projected fermion operators $\tilde{c}_{\alpha,i}^{\dagger}$ and $\tilde{c}_{i}^{\alpha}$ 
\eqref{eqn:AF-Jk-projected-fermion} 
with the fermionic generators \eqref{eqn:suN1-fermionic-gen} of SU($N|1$).  
The idea is to combine the $N$ states ($|\boldsymbol{N};\alpha\rangle$) with a ${\tiny \yng(1)}$-spin occupying the site  
and the Kondo singlet ($|\bullet\rangle$) into a single $(N+1)$-component multiplet.  
Looking at the structure of the SU($N|1$) representation \eqref{eqn:SUN1-irreps-FB}, we see that 
the $n=1$ case of the F-B construction \eqref{eqn:suMN-gen-Abrikosov-1}-\eqref{eqn:suMN-gen-Abrikosov-2} works.  
Specifically, we identify the states $|\boldsymbol{N}; \alpha\rangle$ and the Kondo singlet $|\bullet \rangle$ 
with the states with the boson-fermion occupation $(n_{\text{F}}, n_{\text{B}})=(1,0)$  
and $(n_{\text{F}}, n_{\text{B}})=(0,1)$, respectively  
(see Table~\ref{tab:suN1-corresp-AF-Jk}; this seems quite natural from the original spirit 
of the slave-boson construction \cite{Coleman-84}).  
The commutation relations \eqref{eqn:suM1-CCR-3} suggest us to identify 
\[
\mathcal{Q}_{\alpha} \leftrightarrow \tilde{c}_{\alpha}^{\dagger} \, , \quad 
\widetilde{\mathcal{Q}}_{\alpha} \leftrightarrow \tilde{c}^{\alpha}  \quad (\alpha=1,\ldots,N) 
\]
up to an overall factor.      
If we assign the states $|\boldsymbol{N}; \alpha \rangle$ ($\alpha=1,\ldots,N$) to the first $N$ components, 
and the Kondo singlet $|\bullet \rangle$ to the $(N+1)$-th component [note that the SU($N$)-singlet $|\bullet\rangle$ 
here is not necessarily the same as the {\em physical} Kondo singlet $|\bullet\rangle^{\prime}$], 
the anti-commutator \eqref{eqn:suM1-CCR-2} reads as [use \eqref{eqn:suN1-ACR-by-matrices}]: 
\begin{equation}
 \{ \mathcal{Q}_{\alpha} , \widetilde{\mathcal{Q}}_{\beta} \}  
 = 
|\boldsymbol{N}; \alpha \rangle \langle\boldsymbol{N}; \beta |  
+   \delta^{\beta}{}_{\alpha} |\bullet \rangle \langle\bullet |    \; .
\label{eqn:Q-Q-ACR-by-Hubbard}
\end{equation}
If we identify the many-body basis states
\[ | \cdots \rangle \otimes |\boldsymbol{N};\alpha_{i_{1}} \rangle_{i_{1}} \otimes |\cdots \rangle 
\otimes |\boldsymbol{N};\alpha_{i_{2}} \rangle_{i_{2}} 
\otimes  \cdots 
\]
(where $|\cdots\rangle$ stands for the product of the physical Kondo singlets $|\bullet \rangle_{i}^{\prime}$) 
 in Eqs.~\eqref{eqn:SUN-projected-hopping-1D-1}, 
\eqref{eqn:SUN-projected-hopping-1D-2}, and \eqref{eqn:Kondo-singlet-modified} with 
the states 
\[
\mathcal{Q}_{\alpha_{i_{1}}, i_{1}} \mathcal{Q}_{\alpha_{i_{2}}, i_{2}} \cdots  | \bullet \rangle^{\otimes L} \; ,
\]
all the sign factors arising from the anti-commutation of the original fermions $c_{\alpha,i}^{\dagger}$ 
and $c_{i}^{\alpha}$ are taken into account by that of $\mathcal{Q}_{\alpha,i}$ and $\widetilde{\mathcal{Q}}_{\alpha,i}$ 
on different sites.  
With this identification, we see that the fermionic generators $\{ \mathcal{Q}_{\alpha,i}, \widetilde{\mathcal{Q}}_{\alpha,i} \}$ 
are related to the projected fermion operators as:
\begin{equation}
\mathcal{Q}_{\alpha,i} = (-1)^{N-1} \sqrt{N} \, \tilde{c}_{\alpha,i}^{\dagger}  \, , \quad 
\widetilde{\mathcal{Q}}_{\alpha,i} = (-1)^{N-1} \sqrt{N} \, \tilde{c}_{i}^{\alpha}   
  \; ,
\label{eqn:AF-Jk-c-by-Q}
\end{equation}
which correctly reproduces \eqref{eqn:AF-Jk-projected-ACR} from \eqref{eqn:Q-Q-ACR-by-Hubbard}.   
The bosonic generator $\mathcal{K}_{i}$ essentially counts the number of the Kondo singlets at site-$i$ 
[multiplied by a factor $(N-1)$] which physically corresponds to that of holes:
\begin{equation}
\begin{split}
\mathcal{K}_{i} & = 1 + (N-1) |\bullet \rangle \langle\bullet |_{i}
= 1 + (N-1)(N - \tilde{n}_{i})  \\
& = (N^{2} -N +1) - (N-1) \tilde{n}_{i}  \quad (\tilde{n}_{i} =N-1,N) \; .
\end{split}
\label{eqn:AF-Jk-K-by-fermion}
\end{equation}
The bosonic SU($N$) spin operators $\mathcal{S}^{A}$ are given in Eq.~\eqref{eqn:suN1-bosonic-gen}.  

Using the relations \eqref{eqn:AF-Jk-c-by-Q} and \eqref{eqn:AF-Jk-K-by-fermion}, 
we see that the following effective Hamiltonian at the special point $\mathcal{J} = t/N$ is SU($N|1$)-symmetric: 
\begin{equation}
\begin{split}
\mathcal{H}_{\text{SUSY}}^{\text{(AF)}} & = 
- t \sum_{i} \sum_{\alpha=1}^{N}  \left( 
\tilde{c}_{\alpha,i}^{\dagger} \tilde{c}_{\alpha,i+1} 
+ \tilde{c}_{\alpha,i+1}^{\dagger} \tilde{c}_{\alpha,i} 
\right)  \\
& \phantom{=} 
+ \mathcal{J} \sum_{i} \left\{ 
\sum_{A=1}^{N^{2}-1} \mathcal{S}^{A}_{i} \mathcal{S}^{A}_{i+1} 
- \frac{N-1}{N} \tilde{n}_{i} \tilde{n}_{i+1}  
\right\}  \\
& \phantom{=}
+ \frac{2(N^{2} -N +1)}{N^{2}} t \sum_{i} \tilde{n}_{i} 
+ \text{const.}
 \; .
 \end{split}
 \label{eqn:suN1-sym-interaction-AF}
\end{equation}
The two terms in the second line do not exist in the effective Hamiltonian 
of the usual SU($N$) KLM \eqref{eqn:SUN-KLM-2}.  
In fact, the local SU($N$) spin $S^{A}$ projected onto the ground-state subspace spanned by the states 
\eqref{eqn:basis-Kondo-singlet} and \eqref{eqn:basis-N-rep-AF-Jk} is either $0$ (when the site is occupied by 
the Kondo singlet) or $S^{A}({\tiny \yng(1)}) = G^{A}$ (when the site is in ${\tiny \yng(1)}$).  
This perfectly fits the form of the SU($N | 1$) generators $\mathcal{S}^{A}$ in Eq.~\eqref{eqn:suN1-bosonic-gen}:
\begin{equation}
S^{A} \xrightarrow{\text{proj.}}
\widetilde{S}^{A} = 
\left( 
\begin{array}{c|c}
S^{A}({\tiny \yng(1)}) & 0 \\
\hline 
0 & 0 
\end{array}
\right) 
= \mathcal{S}^{A} \; .
\label{eqn:local-moment-projected-AF}
\end{equation}
Therefore, the first term is obtained just by projecting the Heisenberg interaction among the local spins:  
\[
J_{\text{H}} \, S^{A}_{i} S^{A}_{j}  
\xrightarrow{\text{proj.}}  J_{\text{H}} \, \mathcal{S}^{A}_{i} \mathcal{S}^{A}_{j}  
 \; .
\] 
The second is provided by an attractive interaction between the fermions at sites $i$ and $j$.  
Summarizing all these, we conclude that the supersymmetric interaction \eqref{eqn:suN1-sym-interaction-AF} 
is obtained in the $J_{\text{K}}=\infty$ limit of the (generalized) Kondo-Heisenberg model \eqref{eqn:SUN-KHM}:
\begin{equation*}
\begin{split}
& \mathcal{H}_{\text{KHM}}  \\
& = 
 - t \sum_{i} \sum_{\alpha=1}^{N} 
   \left(c_{\alpha,\,i}^\dag c_{\alpha,\,i+1}  + \text{h.c.}\right)   
+ J_{\text{K}}  \sum_{i}\left( 
\sum_{A=1}^{N^{2}-1}\hat{s}_{i}^{A} S_{i}^{A}
\right) \\
& \phantom{=} 
+ J_{\text{H}}  \sum_{i}\left( 
\sum_{A=1}^{N^{2}-1} S_{i}^{A} S_{i+1}^{A}
\right) 
+ V \sum_{i} n_{i} n_{i+1}
\end{split}
\end{equation*}
with 
\begin{equation}
J_{\text{H}}= \frac{t}{N}\, , \quad 
V = - \frac{N-1}{N^{2}} t  \; .
\end{equation}

A few remarks are in order about the supersymmetric point.  
By construction, it is obvious that the supersymmetric interaction \eqref{eqn:gen-SUN1-Hamiltonian} is defined with respect, not to the physical operators 
(e.g., $c_{\alpha}^{\dagger}$, $c^{\alpha}$) but to the supersymmetry (SUSY) generators 
$\{ S^{A}, \mathcal{K}, \mathcal{Q}, \widetilde{\mathcal{Q}} \}$.  
Therefore, depending on how we identify the SU($N | 1$) generators with the physical operators (and how we define the local degrees of freedom), 
the resulting supersymmetric models may be different.   
For instance, if we realize the SU($N$) $t$-$J$ model \eqref{eqn:suN1-sym-interaction-AF} in the large-$U$ limit of the SU($N$) Hubbard model
 in which multiply-occupied sites are projected out, we have different relations $\mathcal{Q}_{\alpha} = \tilde{c}_{\alpha}^{\dagger}$ and  
$\widetilde{\mathcal{Q}}_{\alpha} =  \tilde{c}^{\alpha}$ instead of \eqref{eqn:AF-Jk-c-by-Q}.   
Plugging these into  Eq.~\eqref{eqn:gen-SUN1-Hamiltonian}, we see that now $\mathcal{J} = t$ 
corresponds to the supersymmetric point \footnote{%
If we normalize the SU($N$) generators as $\text{Tr}(S^{A}S^{B})=\delta^{AB}/2$, the condition reads as: $\mathcal{J} =2t$ 
which is well-known in the SU(2) literature.}.    
On the other hand, the exchange interaction is given by $\mathcal{J} = 2 t^{2}/U$ which must be much smaller than $t$.   
In this sense, the supersymmetric model (derived from the large-$U$ Hubbard model) with $\mathcal{J} =t$  
seems unrealistic.  
However, when we use the large-$J_{\text{K}}$ limit of the SU($N$) Kondo-Heisenberg model \eqref{eqn:SUN-KHM} 
to realize the same SU($N$) $t$-$J$ model, the supersymmetric point corresponds to 
$\mathcal{J} = J_{\text{H}}= \frac{t}{N}\, (< t)$, which seems more feasible.   
\begin{table}[hbt]
\caption{\label{tab:suN1-corresp-AF-Jk} Interpretation of the SU($N$) and SUSY states in terms of fermionic states. 
In the ground-state subspace, the projected fermion number $\tilde{n}$ takes $N-1$ and $N$.}
\begin{ruledtabular}
\begin{tabular}{lccc}
fermionic states  & $\tilde{n}$ & SU($N$) irreps. & $(n_{\text{F}},n_{\text{B}})$  \\
\hline
$ |\text{f}\rangle_{\text{F},i}$ & $N$ & $|\boldsymbol{N}; \alpha\rangle_{i}$ (${\tiny \yng(1)}$) & $(1,0)$ \\
\hline
$|{}^{\alpha}\rangle_{\text{F},i}=c_{i}^{\alpha} |\text{f}\rangle_{\text{F},i} $ & $N-1$ & $|\bullet \rangle_{i}$ (Kondo singlet) 
& $(0,1)$ \\
\end{tabular}
\end{ruledtabular}
\end{table}

\subsection{Ferromagnetic Kondo coupling}
\label{sec:SUSY-FM-JK}
When $J_{\text{K}}$ is ferromagnetically large, the effective Hamiltonian contains the two types of mobile spins 
${\tiny \yng(2)}$ and ${\tiny \yng(1)}$ instead of ${\tiny \yng(1)}$ and the Kondo singlets $\bullet$ for $J_{\text{K}}>0$.  
Accordingly, the local SU($N$) spin operators are given either by $S^{A}({\tiny \yng(1)})$ or by $S^{A}({\tiny \yng(2)})$.   
To describe these two states on an equal footing, we now use SU($N|1$) in the ``B-F'' (or, slave-fermion) construction 
\eqref{eqn:suMN-gen-Schwinger-1} and \eqref{eqn:suMN-gen-Schwinger-2} 
with the total particle number $n=n_{\text{B}}+n_{\text{F}}=2$, in which the boson-fermion occupation 
$(n_{\text{B}}, n_{\text{F}})=(2,0)$ and $(n_{\text{B}}, n_{\text{F}})=(1,1)$ correspond to 
the states ${\tiny \yng(2)}$ and ${\tiny \yng(1)}$, respectively [see Eq.~\eqref{eqn:SUN1-irreps-BF}].  
The correspondence among the (projected) fermion number $\tilde{n}$, the SU($N$) representations, and the SU($N|1$) states 
is summarized in Table~\ref{tab:suN1-corresp-FM-Jk}.   

As in the previous section, we begin with identifying the projected fermion operators \eqref{eqn:FM-Jk-projected-fermion} with 
the fermionic generators $\mathcal{Q}$ and $\widetilde{\mathcal{Q}}$ in the $n=2$ representation 
which is {\em different} from the one used in the previous section.  
Using \eqref{eqn:B-F-n2-rep-basis-1} and \eqref{eqn:B-F-n2-rep-basis-2}, we can write the matrix elements of 
the fermionic generators that create the ${\tiny \yng(2)}$ spin out of ${\tiny \yng(1)}$ as:
\begin{equation}
\begin{split}
 \mathcal{Q}_{\alpha} = & \sqrt{2}\, |(\alpha,\alpha)\rangle \langle \boldsymbol{N}; \alpha |  \\
& + \sum_{\beta > \alpha} |(\alpha,\beta)\rangle \langle \boldsymbol{N}; \beta | 
+ \sum_{\beta < \alpha} |(\beta,\alpha)\rangle \langle \boldsymbol{N}; \beta |  \\
= & \widetilde{\mathcal{Q}}^{\dagger}_{\alpha}   \; ,
\end{split}
\label{eqn:suN1-gen-n2-rep}
\end{equation}
which immediately enables us to identify:
\begin{subequations}
\begin{equation}
 \mathcal{Q}_{\alpha,i} = \sqrt{2} \, \tilde{c}_{\alpha,i}^{\dagger}  \, , \quad 
\widetilde{\mathcal{Q}}_{\alpha,i} = \sqrt{2} \, \tilde{c}_{i}^{\alpha}  \; .
 \end{equation}
 This is natural since $c_{\alpha,i}^{\dagger}$ ($c_{i}^{\alpha}$) creates (annihilates) 
 the ${\tiny \yng(2)}$-particles.  
 The two bosonic generators are now given in terms of the physical operators by:
 \begin{equation}
\mathcal{S}_{i}^{A} = 
\left( 
\begin{array}{c|c}
S_{i}^{A}({\tiny \yng(2)}) & 0 \\
\hline
0 & S_{i}^{A}({\tiny \yng(1)}) 
\end{array}
\right) 
\label{eqn:SUN1-spin-FM}
\end{equation}
[with $S_{i}^{A}({\tiny \yng(2)})$ being the SU($N$) generator 
in the $[N(N+1)/2]$-dimensional representation ${\tiny \yng(2)}$] and 
 \begin{equation}
 \begin{split}
\mathcal{K}_{i}  &= n_{\text{B}} + N n_{\text{F}} 
 = (N+1) - (N - 1)  \tilde{n}_{i} \\
  & = 
 \left( 
\begin{array}{c|c}
2 \times \mathbf{1}_{N(N+1)/2} & 0 \\
\hline
0 & (N+1) \times \mathbf{1}_{N}
\end{array}
\right) 
\end{split}
\end{equation}
\end{subequations}
where we have used $n_{\text{B}}+n_{\text{F}}=2$ and $n_{\text{B}} = \tilde{n}_{i}+ 1$ (with the local 
$c$-fermion number $\tilde{n}_{i} =0,1$).   
Out of these generators, we can readily construct the supersymmetric Hamiltonian \eqref{eqn:gen-SUN1-Hamiltonian}:  
\begin{equation}
\begin{split}
\mathcal{H}_{\text{SUSY}}^{\text{(FM)}}  = &  
 - t \sum_{i} \sum_{\alpha=1}^{N}  \left( 
\tilde{c}_{\alpha,i}^{\dagger} \tilde{c}_{i+1}^{\alpha}
+ \tilde{c}_{\alpha,i+1}^{\dagger} \tilde{c}_{i}^{\alpha} 
\right)  \\
& + \mathcal{J} \sum_{i} \left\{ 
\sum_{A=1}^{N^{2}-1} \mathcal{S}^{A}_{i} \mathcal{S}^{A}_{i+1} 
- \frac{N-1}{N} \tilde{n}_{i} \tilde{n}_{i+1} 
\right\}  \\
& + \frac{N+1}{N} t  \sum_{i} \tilde{n}_{i}  + \text{const.} 
\end{split}
\label{eqn:suN1-sym-interaction-FM}
\end{equation}
with $\mathcal{J}=t/2$.  
This looks the same as \eqref{eqn:suN1-sym-interaction-AF} except that now the particles created or annihilated 
by $\tilde{c}_{\alpha,i}^{\dagger}$ and $\tilde{c}_{i}^{\alpha}$ are ${\tiny \yng(2)}$ spins embedded in 
the ${\tiny \yng(1)}$ background [see Fig.~\ref{fig:SUN-large-Jk-GS-incom}(a-2)].  
Correspondingly, the supersymmetric point {\em in the physical model} is shifted: 
$\mathcal{J}=t/N \,(\text{AF}) \to t/2 \, (\text{FM})$.   
Interestingly, the two supersymmetric models \eqref{eqn:suN1-sym-interaction-AF} and \eqref{eqn:suN1-sym-interaction-FM} 
emerge from the same Kondo-Heisenberg model depending on the sign of $J_{\text{K}}$ and filling ($f=1/N$ or $1-1/N$).  

The origin of the interactions other than the spin exchange is obvious.  
One may think that, as in the antiferromagnetic case, the spin exchange $\mathcal{S}^{A}_{i} \mathcal{S}^{A}_{i+1}$  
comes from the spin-spin interaction ($J_{\text{H}}$) among the local spins.  
However, as the local SU($N$) spin $S^{A}$ acts differently on the two local degrees 
of freedom (${\tiny \yng(2)}$ and ${\tiny \yng(1)}$), 
its projected expression is now given by [see Eq.~\eqref{eqn:local-moment-projected-AF}]:
\begin{equation}
\widetilde{S}_{i}^{A} = 
\left( 
\begin{array}{c|c}
\frac{1}{2} S_{i}^{A}({\tiny \yng(2)}) & 0 \\
\hline
0 & S_{i}^{A}({\tiny \yng(1)})   
\end{array}
\right) 
\end{equation}
which is different from the generator $\mathcal{S}_{i}^{A}$  
appearing in \eqref{eqn:suN1-sym-interaction-FM} [see Eq.~\eqref{eqn:SUN1-spin-FM} for the definition of $\mathcal{S}_{i}^{A}$].  
Therefore, to realize the supersymmetric interaction, we need to slightly modify the $J_{\text{H}}$ interaction 
in the Kondo-Heisenberg model \eqref{eqn:SUN-KHM} in such a way that it includes diagonal (i.e. off-site) 
Kondo couplings as well:
\begin{equation}
\begin{split}
& J_{\text{H}}  \sum_{i}\left( 
\sum_{A=1}^{N^{2}-1} S_{i}^{A} S_{i+1}^{A}
\right)   \\
&  \to 
J_{\text{H}}  \sum_{i}\left\{ 
\sum_{A=1}^{N^{2}-1} (\hat{s}_{i}^{A} + S_{i}^{A}) (\hat{s}_{i+1}^{A} + S_{i+1}^{A}) 
\right\}    \;  .
\end{split}
\end{equation}
Then, setting  
\begin{equation}
J_{\text{H}}= \frac{t}{2}\, , \quad 
V = - \frac{N-1}{2N} t  
\end{equation}
and $J_{\text{K}}= -\infty$ in the generalized Kondo-Heisenberg model \eqref{eqn:SUN-KHM} realizes 
the supersymmetric model \eqref{eqn:suN1-sym-interaction-FM}.  

Unfortunately, the behavior of the ``higher-spin'' SU($N$) $t$-$J$ model \eqref{eqn:suN1-sym-interaction-FM} 
is not known except at $\mathcal{J}=0$ where we have rigorously established in Sec.~\ref{sec:ferro-FM-Kondo} 
that the ground state is ferromagnetic.  
The inclusion of the $\mathcal{J}\, (>0)$-term that favors antiferromagnetic correlation may 
destabilize the ferromagnetic ground state as in the $N=2$ case \cite{Sikkema-A-W-97,Masui-T-22}.   
In this sense, the supersymmetric point at which the two tendencies compete might play a special role 
and the search for the exactly solvable supersymmetric ``spin'' Hamiltonians \cite{Arovas-H-Q-Z-09,Hasebe-T-11,Hasebe-T-13} 
would be interesting.  
\begin{table}[hbt]
\caption{\label{tab:suN1-corresp-FM-Jk} Correspondence among fermionic states, SU($N$)-spin, and 
SUSY representation}
\begin{ruledtabular}
\begin{tabular}{lccc}
fermionic states  & $\tilde{n}$ & SU($N$) spin states  & $(n_{\text{B}},n_{\text{F}})$   \\
\hline
$|{}_{\alpha}\rangle_{\text{F},i}=c^{\dagger}_{\alpha,i} | 0\rangle_{\text{F},i} $ & $1$ 
& $| (\alpha,\beta) \rangle_{i}$ (${\tiny \yng(2)}$)
& $(2,0)$ \\
\hline
$ | 0\rangle_{\text{F},i}$ & $0$ & $|\boldsymbol{N}; \alpha\rangle_{i}$ (${\tiny \yng(1)}$) & $(1,1)$ \\
\end{tabular}
\end{ruledtabular}
\end{table}

\section{Summary and discussion}
\label{sec:conclusion}
\begin{figure}[htb]
\begin{center}
\includegraphics[width=\columnwidth,clip]{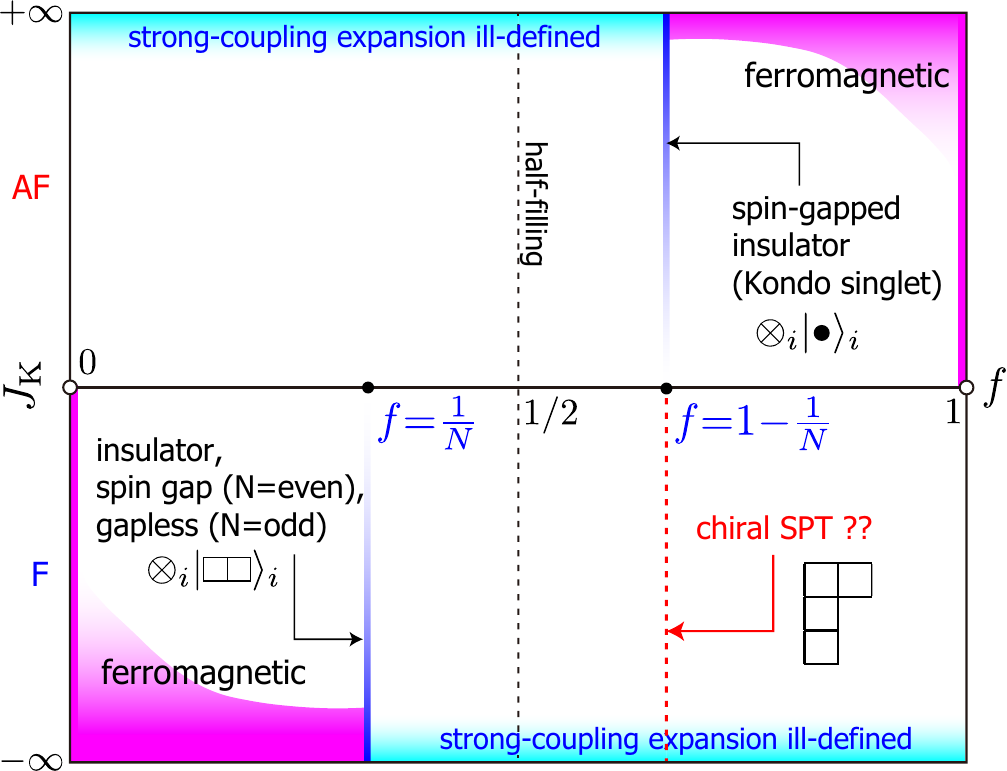}
\end{center}
\caption{Strong-coupling phases in SU($N$) Kondo lattice model in 1D.  
The insulating phase at $f=1/N$ ($J_{\text{K}}<0$) has a finite spin gap only when $N$ is even, while the one at $f=1-1/N$ 
($J_{\text{K}}>0$) is fully gapped.   The ferromagnetic phases (highlighted in magenta) 
extend to the lower-density (when $J_{\text{K}}<0$) 
or higher-density ($J_{\text{K}}>0$) side of these insulating phases.  At the extreme limits $f\to 0$ ($\mathcal{N}_{\text{c}}=1$) 
or $f \to 1$ ($\mathcal{N}_{\text{c}}=NL -1$; one hole), 
we can establish ferromagnetism {\em regardless} of the magnitude of $J_{\text{K}}(\neq 0)$.   
\label{fig:SUN-KLM-phase-diag}
}
\end{figure}

In this paper, we have considered the ground state of the SU($N$) Kondo lattice model 
with the local spins in the $N$-dimensional defining representation (${\tiny \yng(1)}$)  
for sufficiently strong Kondo coupling.  
Specifically, we have shown rigorously that the ground state of the one-dimensional model 
(with open boundary condition) for fillings $0 < f  < 1/N$ 
(when $J_{\text{K}} <0$) and $1-1/N < f < 1$ (when $J_{\text{K}} >0$) is ferromagnetic.  
The corresponding ferromagnetic states are shown schematically in Figs.~\ref{fig:SUN-ferro-config}(b) and (c).  
The proof is based on the Perron-Frobenius theorem on the spectral properties of irreducible non-positive matrices.  
In higher dimensions, we can make a similar statement on ferromagnetism, e.g.,   
when there is precisely one fewer fermion from the commensurate filling $f=1/N$ or $1$.  
We can also treat the problem with one fermion or one hole exactly for any $J_{\text{K}} \neq 0$ 
(i.e., without relying on the strong-coupling limit) to prove the ferromagnetic ground states.  
Therefore, in the extreme limits $f \to 0$ ($J_{\text{K}}<0$) 
and $f \to 1$ ($J_{\text{K}}>0$), the ferromagnetic phases are expected to persist down to $J_{\text{K}} \to \pm 0$ 
(see Fig.~\ref{fig:SUN-KLM-phase-diag}).  
For the situation considered in this paper, we can generalize the double-exchange scenario to SU($N$) with 
due complication, which semi-classically explains the occurrence of ferromagnetism for $J_{\text{K}} <0$.   
Considering the large positive scattering $g$-$e$ length of ${}^{173}\text{Yb}$ which suggests a large ferromagnetic 
$J_{\text{K}}$, ${}^{173}\text{Yb}$ would be a promising system to test the SU($N$) double-exchange mechanism 
in the strongly-coupled KLM.  

At the special filling fractions $f=1/N$ and $1-1/N$, the system is insulating.   
When $f=1-1/N$ and $J_{\text{K}} \gg t$, the system is a spin-gapped insulator, which is the SU($N$)-analog of the well-known 
Kondo insulator in the usual SU(2) KLM at half-filling.  
On the other hand, in the insulating phase at $f=1/N$ and $- J_{\text{K}} \gg t$,  
the behavior of the spin sector depends on the parity of $N$; the spin correlation is algebraic (exponentially decaying) 
when $N=\text{odd}$ ($N=\text{even}$).      
The schematic phase diagram that summarizes the main results of this paper is given in Fig.~\ref{fig:SUN-KLM-phase-diag}.  

As has been seen in Sec.~\ref{sec:strong-coupling}, 
the low-energy physics at strong coupling is described by two different degrees of freedom [Kondo singlets and mobile 
$\boldsymbol{N}$ spins for $J_{\text{K}}>0$, and mobile two species of spins $\boldsymbol{N}$ (${\tiny \yng(1)}$) 
and ${\tiny \yng(2)}$ for $J_{\text{K}}<0$].  
We have found that the language of super Lie algebra SU($N | 1$) perfectly fits into these situations and can describe the low-energy processes in a natural way.  In fact, for particular sets of 
parameters, we can realize the supersymmetric models in the limit of strong Kondo coupling.   
The resulting conditions for supersymmetry are milder (or more realistic) than the one known for the $t$-$J$ model.  
\section*{Acknowledgements}
The authors would like to thank S.~Capponi, K.~Hasebe, H.~Katsura, N.~Kawakami, P.~Lecheminant, 
K.~Ono, D.~Papular, and Y.~Takahashi for helpful discussions 
and correspondences. 
The author is supported in part by the Japan Society for the Promotion of Science (JSPS) KAKENHI Grant No. 18K03455 and No. 21K03401.  

\appendix
\section{Young diagrams and SU($N$) representations}
\label{sec:Young-diag}
In this appendix, we give a quick explanation of what the Young diagrams stand for in physical terms. 
Let us first introduce the fundamental representations that are building blocks of all possible irreducible representations.  
There are $N-1$ fundamental representations $\mathcal{R}_{n}$ 
each of which is realized by a fixed number $n(=1,\ldots, N-1)$ 
of $N$-colored fermions $c^{\dagger}_{\alpha}$ ($\alpha=1,\ldots, N$) 
[the two cases $n=0,N$ correspond to SU($N$)-singlet and are trivial].  
The $n$-fermion representation $\mathcal{R}_{n}$ is spanned by the states of the form 
(the bracket $[\cdots]$ stands for anti-symmetrization):
\begin{equation}
\begin{split} 
& |{}_{[ \alpha_{1}, \ldots , \alpha_{n} ] } \rangle := 
c^{\dagger}_{\alpha_{1}} c^{\dagger}_{\alpha_{2}} \cdots c^{\dagger}_{\alpha_{n}} |0\rangle_{\text{F}} 
\end{split}
\end{equation}
and has dimensions $\frac{N!}{(N-n)!n!}$.   If necessary, we can easily calculate the corresponding matrix representation 
using the second quantized generators similar to \eqref{eqn:second-quantized-gen}.   
We assign the following single-column 
Young diagrams
\begin{equation}
\mathcal{R}_{n}: \quad 
\text{\scriptsize $n$} \left\{ 
{\tiny \yng(1,1,1,1)  }
\right.  
\quad (n=1,\ldots, N-1)  
\end{equation}
to these representations.  By construction, the $n$ boxes in the same column are anti-symmetrized.  
The simplest of them is the $N$-dimensional representation ${\tiny \yng(1) }$ 
which is spanned by the following $N$ single-fermion ($n=1$) states:
\[
|{}_{\alpha}\rangle := c^{\dagger}_{\alpha} |0\rangle_{\text{F}}  \quad (\alpha=1,\ldots, N)  
\]
and has been used for the local spins of the models \eqref{eqn:SUN-KLM-2} and \eqref{eqn:SUN-KHM}.  

The conjugate representation $\overline{\mathcal{R}}_{n}$ of $\mathcal{R}_{n}$ is obtained by applying the particle-hole 
transformation:
\begin{equation*}
\begin{split}
& |{}^{[ \alpha_{1},\ldots , \alpha_{n} ] } \rangle :=  
c^{\alpha_{n}} \cdots c^{\alpha_{1}} |\text{f}\rangle_{\text{F}}  \\
& = \frac{1}{(N-n)!} \sum_{\{ \beta_{i} \}} \epsilon^{\alpha_{1}\cdots \alpha_{n} \beta_{n+1}\cdots \beta_{N}}   
| \underbrace{ {}_{[ \beta_{n+1} ,\cdots , \beta_{N}] }}_{N-n}  \rangle  \\
& \left( |\text{f}\rangle_{\text{F}} = c^{\dagger}_{1} \cdots c^{\dagger}_{N} |0\rangle_{\text{F}}   \right) \; .
\end{split}
 \end{equation*}
 As the right-hand side transforms like $\mathcal{R}_{N {-}n}$, the conjugation transforms the Young diagram as: 
 \begin{equation}
\text{\scriptsize $n$} \left\{ 
{\tiny \yng(1,1,1,1)  }
\right.  \; (\mathcal{R}_{n} )  \; \xrightarrow{\text{conjugate}} \; 
\text{\scriptsize $N {-} n$} \left\{ 
{\tiny \yng(1,1)  }
\right.   \; (\overline{\mathcal{R}}_{n}  = \mathcal{R}_{N {-}n} )
\;   .
\label{eqn:conjugation-fundamental-rep}
\end{equation}
Clearly, the $N$ one-hole states 
\[
|{}^{\alpha} \rangle =  c^{\alpha} |\text{f}\rangle_{\text{F}} 
= (-1)^{\alpha-1} \prod_{\beta \neq \alpha} c_{\beta}^{\dagger} |0\rangle_{\text{F}}
\quad (\alpha=1,\ldots, N)
\]
appearing in Eq.~\eqref{eqn:2-low-energy-states-high-density} span the conjugate $\overline{\mathcal{R}}_{1}$ 
of the one-fermion representation $\mathcal{R}_{1}$ (${\tiny \yng(1)}$).   

\begin{figure}[hbt]
\begin{center}
\includegraphics[scale=0.4]{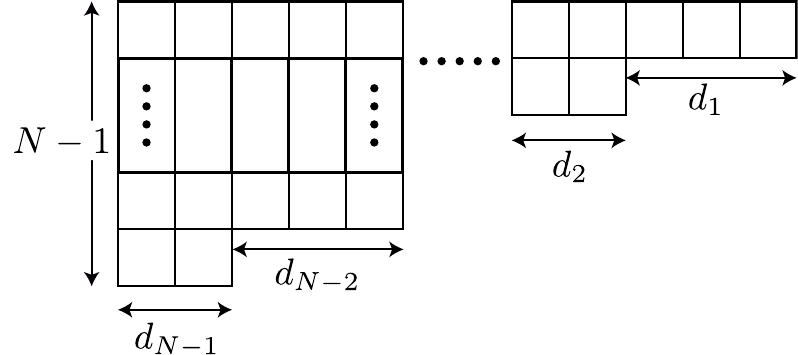}
\end{center}
\caption{The Young diagram corresponding to the SU($N$) irreducible representation specified by the Dynkin labels 
$(d_1,d_2,\ldots, d_{N-1})$.
\label{fig:Dynkin-Young}}
\end{figure}

The generic irreducible representations are constructed by tensoring the $N-1$ fundamental representations 
$\mathcal{R}_{n}$:
\begin{equation}
\mathcal{R}_{1}^{\otimes d_{1}} \otimes \cdots \otimes \mathcal{R}_{N-1}^{\otimes d_{N-1}} \; .
\end{equation}
In doing so with fermions, we need to introduce an additional degree of freedom (``flavor'') on top of the color 
$\alpha(=1,\ldots,N$).   
The set of non-negative integers (Dynkin labels) $( d_{1}, \ldots, d_{N-1} )$ uniquely specifies the irreducible representation.  
The Young diagram corresponding to a generic representation $( d_{1}, \ldots, d_{N-1} )$ is made of 
$d_{1}$ length-1 columns, $d_{2}$ length-2 ones, and so on (see Fig.~\ref{fig:Dynkin-Young}).  
In SU(2), only the representations of the form 
\[   \underbrace{\tiny \yng(5)}_{2S} \]
are allowed and the number of boxes $d_{1} = 2S$ suffices to label them.  

For example, the diagram
\[
{\tiny \yng(3,1)} 
\]
stands for the representation $(2,1,0,\ldots,0)$, while the adjoint representation $(1,0,\ldots, 0,1)$ 
under which the SU($N$) generators transform is specified as:
\begin{equation}
\text{\scriptsize $N {-} 1$} \left\{ 
{\tiny
\yng(2,1,1,1,1)
}
\right.  \; .
\label{eqn:Young-adjoint}
\end{equation} 
The conjugate of a given representation is obtained by applying the rule \eqref{eqn:conjugation-fundamental-rep} 
to each column of the corresponding Young diagram and then rearranging the columns into the correct form.  
For instance, the adjoint representation \eqref{eqn:Young-adjoint} is self-conjugate.  
\section{SU($N$) Kondo energy}
\label{sec:Kondo-energy}
In this Appendix, we explicitly calculate the Kondo energy
\[
e_{\text{K}} = J_{\text{K}}
\sum_{A=1}^{N^{2}-1}\hat{s}_{i}^{A} (\mathcal{R}_{\text{F}}) S_{i}^{A} (\mathcal{R}_{\text{S}})   \; 
\]
when the itinerant fermions and the local spin are in the SU($N$) spin states $\mathcal{R}_{\text{F}}$ 
and $\mathcal{R}_{\text{S}}$, respectively.  In this paper, we only consider the case 
with $\mathcal{R}_{\text{S}} = {\tiny \yng(1)}$, and, when there are $n_{\text{c}}$ fermions,  
\[    \mathcal{R}_{\text{F}} = 
\text{\scriptsize $n_{\text{c}}$} \left\{ 
{\tiny \yng(1,1,1) }
\right.   \; . \]
The Kondo energy is conveniently calculated by using the quadratic Casimir  
$\mathcal{C}_{2}(\mathcal{R}) = \sum_{A} S^{A}(\mathcal{R})S^{A}(\mathcal{R})$ 
which is essentially the squared spin:
\begin{equation}
\begin{split}
& \mathcal{C}_{2} (\mathcal{R}_{\text{F}},\mathcal{R}_{\text{S}}) 
= \sum_{A=1}^{N^{2}-1} \left\{ \hat{s}^{A}(\mathcal{R}_{c})  + S^{A} (\mathcal{R}_{\text{S}}) \right\}^{2}   \\
& = \mathcal{C}_{2} (\mathcal{R}_{\text{F}}) + \mathcal{C}_{2} (\mathcal{R}_{S}) 
+ 2  \sum_{A=1}^{N^{2}-1} \hat{s}^{A} (\mathcal{R}_{\text{F}}) S^{A} (\mathcal{R}_{\text{S}})   \; ,
\end{split}
\end{equation}
where $\mathcal{C}_{2}$ is given explicitly by:
\begin{equation}
\mathcal{C}_{2}(\boldsymbol{\lambda}) = 
\sum_{i,j=1}^{N-1}(\mathbf{m} + \mathbf{e})_{i} (\mathbb{K}^{-1})_{ij} 
(\mathbf{m} + \mathbf{e})_{j} 
-  \frac{1}{12}N(N^{2}-1) 
\end{equation}
where the $(N-1)\times(N-1)$ matrix $\mathbb{K}^{-1}$ is the inverse of the Cartan matrix: 
\begin{equation}
(\mathbb{K}^{-1})_{ij} = 
\begin{cases}
\frac{1}{N} i(N - j)  & \text{for } i \leq j \\
\frac{1}{N} (N - i) j  & \text{for } i > j   
\end{cases}
\end{equation} 
and 
\begin{equation*}
\mathbf{e} := \underbrace{%
(1,1,\ldots, 1)
}_{N-1}  \; .
\end{equation*}
The vector $\mathbf{m}$ is the collection of the Dynkin label $m_{i}$ [$\mathbf{m}=(m_{1},\ldots, m_{N-1})$] 
which is the number of length-$i$ columns in the Young diagram.  
For instance, 
\[
 \text{\scriptsize $n_{\text{c}}$} \left\{ 
{\tiny \yng(3,1,1)}
\right.   
\;\; \Leftrightarrow   \;\;  \mathbf{m} = (2,0,\ldots,0,\underset{n_{\text{c}}}{1} ,0,\ldots,0)  \; .
\]
For the fermion density $n_{\text{c}}$ (i.e., $n_{\text{c}}$ fermions per site, or filling $f = n_{\text{c}} /N$), we need the following 
decomposition:
\begin{equation}
\begin{split}
& 
\underbrace{ 
 \text{\scriptsize $n_{\text{c}}$} \left\{ 
\yng(1,1,1)
\right. 
}_{\text{itinerant}} \; 
\otimes 
\underbrace{
\yng(1) 
}_{\text{local spin}}
\;  \sim  \; \;
\text{\scriptsize $n_{\text{c}}{+}1$} \left\{ 
\yng(1,1,1,1) 
\right. 
\;\; \oplus  \;\;
\text{\scriptsize $n_{\text{c}}$} \left\{ 
\yng(2,1,1) 
\right.  \\
&
(1 \leq n_{\text{c}} \leq N-1) \; .
\end{split}
\label{eqn:CG-decomp-SUN-KLM}
\end{equation}
The cases $n_{\text{c}} =0$ and $n_{\text{c}} =N$ correspond respectively 
to the empty and doubly-occupied sites in the usual SU(2) Kondo lattice and are trivial:
\begin{equation}
\underbrace{ \bullet 
}_{\text{itinerant}}
\otimes 
\underbrace{
\yng(1)
}_{\text{local spin}}
\;  \sim  \; \;
\yng(1) 
\quad (n_{\text{c}} =0, N)   \; .
\label{eqn:CG-decomp-SUN-KLM-2}
\end{equation}
The values of the quadratic Casimir for the representations appearing in \eqref{eqn:CG-decomp-SUN-KLM} 
and \eqref{eqn:CG-decomp-SUN-KLM-2} are:
\begin{subequations}
\begin{align}
& \mathcal{C}_{2}({\tiny \yng(1)}) = \frac{1}{N}(N^{2}-1) 
\quad  \text{(defining rep.)}  \\
\begin{split}
& \mathcal{C}_{2}\left({\tiny \text{\scriptsize $n_{\text{c}}{+}1$} \left\{\yng(1,1,1,1)\right. } \, \right)
= \frac{N+1}{N}(n_{\text{c}} +1) \{ N- (n_{\text{c}} +1) \} \\ 
&  \text{[anti-symmetric $(n_{\text{c}}+1)$-tensor]}   
\end{split}
\\
\begin{split}
& \mathcal{C}_{2}\left({\tiny \text{\scriptsize $n_{\text{c}}$} \left\{\yng(2,1,1)\right. } \, \right) \\
& = \frac{n_{\text{c}} +1}{N}\left\{ N(N-(n_{\text{c}} +1)) + (3N - (n_{\text{c}} +1)) \right\}   \\
& (1 \leq n_{\text{c}} \leq N-1)  \;  .
\end{split}
\end{align}
\end{subequations}
From these, the Kondo energy \eqref{eqn:def-Kondo-energy} is readily calculated as:
\begin{equation}
\begin{split}
& e_{\text{K}} = -\frac{N+1}{N} n_{\text{c}} J_{\text{K}} \quad \text{for } \;\;  
{\tiny \text{\scriptsize $n_{\text{c}}{+}1$} \left\{ 
\yng(1,1,1,1) \right.   } 
\; , \\
& e_{\text{K}} = \left( 1-\frac{n_{\text{c}}}{N} \right) J_{\text{K}} \quad \text{for } \;\;  
{\tiny   \text{\scriptsize $n_{\text{c}}$} \left\{ 
\yng(2,1,1)
\right. }
\quad (1 \leq n_{\text{c}} \leq N-1)    \\
& e_{\text{K}} = 0 \quad (n_{\text{c}}=0,N) 
\; .
\end{split}
\label{eqn:Kondo-energies-SUN-appendix}
\end{equation}
As is shown in Fig.~\ref{fig:Kondo-energies}, the Kondo energy $e_{\text{K}}$ is concave at 
$n_{\text{c}}=1$ (when $J_{\text{K}} < 0$) or $n_{\text{c}}=N - 1$ (when $J_{\text{K}} > 0$), 
and except there it is linear in $n_{\text{c}}$.     

\section{Irreducibility of the Hamiltonian}
\label{sec:indecomposability}
In this appendix, we sketch the proof of the irreducibility of the one-dimensional effective Hamiltonian 
\eqref{eqn:SUN-projected-hopping-FM-1D}.    
The proof is based on mathematical induction with respect 
to the system size $L(\geq 2)$ \cite{Masui-T-22}.   
Suppose that the Hamiltonian is irreducible for a system sizes $L=L_{0}$, and for {\em all} values of $\mathcal{N}_{\text{c}}$ 
($1 \leq \mathcal{N}_{\text{c}} \leq L_{0}-1$) and any total weights $\Lambda_{\text{tot}}$ allowed for $L_{0}$ 
and $\mathcal{N}_{\text{c}}$.  
Since the Hamiltonian is identically zero 
when $\mathcal{N}_{\text{c}} =0$ (no fermion to move) and when $\mathcal{N}_{\text{c}} =L_{0}$ (no hole to move), 
we must exclude these cases as trivial.  

To find the connectivity structure, we group the basis states of the $(L_{0}+1)$-site system  
(with the total SU($N$) weight $\Lambda_{\text{tot}}$ and the fermion number $1 \leq \mathcal{N}_{\text{c}} \leq L_{0}$) 
according to the states at the site-$(L_{0}+1)$:
\begin{equation}
\begin{split}
\text{(i):} \; \; 
& |\{ \lambda_{i} \} ; \alpha \rangle_{ i_{1},i_{2}, \cdots, i_{\mathcal{N}_{\text{c}} }  }  \\
& =  \left| i_{1},i_{2}, \cdots, i_{\mathcal{N}_{\text{c}} } ;  \{ \lambda_{i} \}_{\sum_{i=1}^{L_{0}} \lambda_{i} 
= \Lambda_{\text{tot}} - \lambda_{\alpha}} \right\rangle \otimes | {\tiny \yng(1)}; \lambda_{\alpha} \rangle_{L_{0}+1}  \\
& \quad ( \alpha=1,\ldots, N)
 \\
\text{(ii):} \; \; 
& |\{ \lambda_{i} \} ; (\alpha,\beta)  \rangle_{ i_{1},i_{2}, \cdots, i_{\mathcal{N}_{\text{c}} -1} , L_{0}+1  }   \\
& =  \left| i_{1},i_{2}, \cdots, i_{\mathcal{N}_{\text{c}} -1} ;  \{ \lambda_{i} \}_{\sum_{i=1}^{L_{0}}\lambda_{i} 
= \Lambda_{\text{tot}}- \lambda_{\alpha} - \lambda_{\beta}} \right\rangle  \\
& \phantom{ = } 
\otimes \left| {\tiny \yng(2)} ;  \widetilde{\lambda}_{(\alpha,\beta)} \right\rangle_{L_{0}+1}  
 \quad   ( 1 \leq \alpha \leq \beta \leq N )  \; ,
\end{split}
\label{eqn:connectivity-basis-SUN}
\end{equation}
where the sequence $\{ i_{1},i_{2}, \cdots, i_{\mathcal{N}_{\text{c}} } \}$ specifies the positions 
of fermions (i.e., those of ${\tiny \yng(2)}$-spins), and  
the set of the local SU($N$) weights $\{ \lambda_{i} \}$ ($i=1,\ldots, L_{0}$) satisfies 
$\sum_{i=1}^{L_{0}} \lambda_{i} + \lambda_{L_{0}+1} = \Lambda_{\text{tot}}$ 
($\lambda_{L_{0}+1} =\lambda_{\alpha} , \,  \widetilde{\lambda}_{(\alpha,\beta)}; \; 
\widetilde{\lambda}_{(\alpha,\beta)}= \lambda_{\alpha} + \lambda_{\beta}$).    
In (i), all the $\mathcal{N}_{\text{c}}$ fermions are contained in the $L_{0}$-site subsystem, while in (ii), one of 
the fermions is sitting at site-$(L_{0}+1)$.  

\begin{figure}[htb]
\begin{center}
\includegraphics[width=\columnwidth,clip]{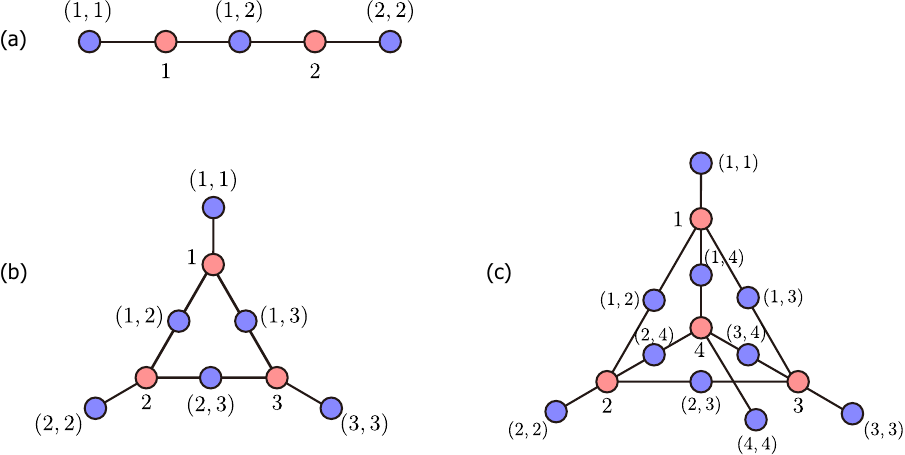}
\end{center}
\caption{Graphical representation of the connectivity introduced by the hopping between 
$L_{0}$ and $(L_{0}+1)$ for (a) $N=2$, (b) $N=3$, and (c) $N=4$.  
The $N(N+3)/2$ groups of the basis states in \eqref{eqn:connectivity-basis-SUN} are represented 
by the vertices (both red and blue). 
By the assumption, all the states {\em within} each group are connected to each other by the $L_{0}$-site Hamiltonian. 
The hopping between the sites $L_{0}$ and $L_{0}+1$ introduces the edges connecting these vertices 
(see the text for how these graphs are drawn).   
\label{fig:connectivity_diagram_SUN}}
\end{figure}

When the hopping between the sites $L_{0}$ and $(L_{0}+1)$ is absent, the effective Hamiltonian assumes 
a block-diagonal form, in which each of the block matrices is irreducible by the assumption except for 
$\mathcal{N}_{\text{c}}=1$ and $L_{0}$ \footnote{%
When $\mathcal{N}_{\text{c}}=1$, $\mathbf{B}_{(\alpha,\beta)}$ 
are not irreducible, while when 
$\mathcal{N}_{\text{c}}=L_{0}$, $\mathbf{B}_{\alpha}$ is not. Therefore, these two cases must be treated separately.}.    
We denote these $N(N+3)/2$ diagonal blocks by: $\mathbf{B}^{{\tiny \yng(1)}}_{\alpha}$ $(\alpha=1,\ldots,N$) 
and $\mathbf{B}^{{\tiny \yng(2)}}_{(\alpha,\beta)}$ $(1 \leq \alpha, \beta \leq N$).   
When the hopping between $L_{0}$ and $(L_{0}+1)$ is switched on, the following transitions are allowed:
\begin{equation}
\begin{split}
&  
|\{ \lambda_{i} \} ; \alpha \rangle_{ i_{1},i_{2}, \cdots, i_{\mathcal{N}_{\text{c}}-1}, i_{\mathcal{N}_{\text{c}}}=L_{0}   }  \\
&  \to  \sum_{\beta=1}^{N}  |\{ \lambda^{\prime}_{i} \} ; (\alpha,\beta)  
    \rangle_{ i_{1},i_{2}, \cdots, i_{\mathcal{N}_{\text{c}} -1} , L_{0}+1  }  \\
& \qquad 
[\lambda^{\prime}_{i}=\lambda_{i} \; (i=1,\ldots, L_{0}-1), \; \lambda^{\prime}_{L_{0}} = \lambda_{L_{0}} - \lambda_{\beta}] \\
& |\{ \lambda_{i} \} ; (\alpha,\beta)  \rangle_{ i_{1},i_{2}, \cdots, i_{\mathcal{N}_{\text{c}} -1} (< L_{0}), L_{0}+1  }    \\
& \to 
 \sum_{\gamma}  |\{ \lambda^{\prime}_{i} \} ; \gamma \rangle_{ i_{1},i_{2}, \cdots, i_{\mathcal{N}_{\text{c}} } =L_{0} }  \\
& (\gamma=\alpha, \beta \text{ when } \alpha\neq \beta, 
 \gamma=\alpha \text{ when } \alpha=\beta; \;
\sum_{i=1}^{L_{0}}\lambda^{\prime}_{i} = \Lambda_{\text{tot}} - \lambda_{\gamma} )\;  ,
\end{split}
\label{eqn:inter-block-tr-SUN}
\end{equation}
where we have omitted the non-zero numerical coefficients. 
The new connectivity structure introduced by the hopping between the sites $L_{0}$ and $(L_{0}+1)$ may be 
best visualized by the graphs shown in Fig.~\ref{fig:connectivity_diagram_SUN}.   
To construct the graph representing the connectivity among the $N(N+3)/2$ groups of basis states, 
we first draw a complete graph made of $N$ vertices (colored in pink) $\alpha=1,\ldots, N$, in which 
each vertex represents one of the type-(i) groups of states in \eqref{eqn:connectivity-basis-SUN} and is associated with 
the block matrices $\mathbf{B}^{{\tiny \yng(1)}}_{\alpha}$.   
By the assumption, all the states contained in the vertex $\alpha$ are connected to each other by 
the action of $\mathbf{B}^{{\tiny \yng(1)}}_{\alpha}$.  
Then, on the edges $(\alpha,\beta)$ ($\alpha < \beta$), we add $N(N-1)/2$ new vertices $(\alpha,\beta)$ (colored in blue) 
corresponding to the type-(ii) basis states in \eqref{eqn:connectivity-basis-SUN}.  Last, we add $N$ new vertices  
$(\alpha,\alpha)$ ($\alpha=1,\ldots, N$) and connect them to the vertices $\alpha$.   
Obviously, all the vertices are connected, which immediately means that any given basis state of the form 
\eqref{eqn:connectivity-basis-SUN} can be transferred to an arbitrary state by the action of the Hamiltonian 
except when $\mathcal{N}_{\text{c}}=1, L_{0}$.     

The two exceptional cases $\mathcal{N}_{\text{c}}=1$ and $L_{0}$ can be handled without relying on the induction.  
In fact, in the above two cases we use strings of (projected) hopping operators to realize SU($N$) ``spin-flips'' 
on a pair of distant sites that allow us to transform any given state to an arbitrary one.  
Therefore, we see that the effective Hamiltonian \eqref{eqn:SUN-projected-hopping-FM-1D} is irreducible 
in the $(L_{0}+1)$-site system as well, which completes the proof.  
\section{Supersymmetry SU(N$|$1)}
\label{sec:SUSY-SU_N1}
In Eq.~\eqref{eqn:AF-Jk-projected-ACR}, we have seen that the fermion operators \eqref{eqn:AF-Jk-projected-fermion} 
projected onto the subspace spanned by the Kondo singlet $ |\bullet \rangle_{i}$ and the fully-occupied state 
$|\boldsymbol{N}; \alpha \rangle_{i}$ do not obey the standard anti-commutation relations.  
In fact, they satisfy the anti-commutation relations of the fermionic generators of the super Lie algebra SU($N|1$) 
(for a quick introduction to SU($N|1$), see appendix A of Ref.~\cite{Hasebe-fuzzy-11}).  

\subsection{Definition}
The superalgebra SU($N | 1$) consists of $N^{2}$ bosonic and $2N$ fermionic generators.   
The bosonic generators are given by the following $(N+1)\times (N+1)$ block-diagonal matrices: 
\begin{equation}
\mathcal{S}^{A} = 
\left( 
\begin{array}{c|c}
G^{A}  & 0 \\ \hline 0 & 0 
\end{array} 
\right)   \;\; (A=1,\ldots, N^{2}-1) , \quad 
 \mathcal{K} = 
 \left( 
\begin{array}{c|c}
 \mathbf{1}_{N}  & 0 \\ \hline 0 & N 
\end{array} 
\right)   
\label{eqn:suN1-bosonic-gen}
\end{equation}
with $G^{A}$ being the SU($N$) generators in the defining representations $\boldsymbol{N}$ which are 
normalized as:
\begin{equation*}
\text{Tr}(G^{A} G^{B} ) = \delta^{AB}   \; .
\end{equation*}
On top of the above $N^{2}$ bosonic generators, there are $2N$ fermionic ones: 
\begin{equation}
\begin{split}
& \mathcal{Q}_{\alpha} := 
\left( 
\begin{array}{c|c}
\mathbf{0}_{N} & \boldsymbol{\tau}_{\alpha} \\ \hline 
\mathbf{0}_{1{\times}N} & 0 
\end{array}
\right) \; , \quad 
\widetilde{\mathcal{Q}}_{\alpha} := 
\left( 
\begin{array}{c|c}
\mathbf{0}_{N} & \mathbf{0}_{N{\times}1} \\ \hline 
(\boldsymbol{\tau}_{\alpha})^{\text{T}} & 0  
\end{array}
\right)   \\
& (\alpha=1,\ldots, N)   \; ,
\end{split}
\label{eqn:suN1-fermionic-gen}
\end{equation}
where the $N$-component column vector $\boldsymbol{\tau}_{\alpha 1}$ has only one non-zero entry:
\begin{equation}
\boldsymbol{\tau}_{\alpha}  = (0, \ldots, 0, \underset{\alpha}{1} ,0 \ldots,0)^{\text{T}}
\end{equation}
and hence $\widetilde{\mathcal{Q}}_{\alpha}=(\mathcal{Q}_{\alpha})^{\text{T}}$ holds.   
Physically, the fermionic generators bring about transitions between the first $N \times N$ block and 
the second one-dimensional one. 

The above generators satisfy the following algebra:
\begin{subequations}
\begin{align}
& [\mathcal{S}^{A} , \mathcal{S}^{B} ]= i f^{ABC} \mathcal{S}^{C}  \, , \quad 
 [\mathcal{K} , \mathcal{S}^{A} ]  = 0  
 \label{eqn:suM1-CCR-1}
\\
\begin{split}
& \{ \mathcal{Q}_{\alpha} , \mathcal{Q}_{\beta} \} = 
 \{ \widetilde{\mathcal{Q}}_{\alpha} , \widetilde{\mathcal{Q}}_{\beta} \} =  0  \\
 & \{ \mathcal{Q}_{\alpha} , \widetilde{\mathcal{Q}}_{\beta} \} = 
[G^{A}]_{\beta\alpha} \mathcal{S}^{A}  + \frac{1}{N} \delta_{\alpha\beta} \mathcal{K} 
 \end{split}
  \label{eqn:suM1-CCR-2}
 \\
\begin{split}
& [ \mathcal{S}^{A} , \mathcal{Q}_{\alpha} ] = \mathcal{Q}_{\beta} [S^{A}]_{\beta\alpha} \, , \\
& [ \mathcal{S}^{A} , \widetilde{\mathcal{Q}}_{\alpha} ] 
= - \widetilde{\mathcal{Q}}_{\beta} [S^{A}]_{\alpha\beta} = \widetilde{\mathcal{Q}}_{\beta} [- (S^{A})^{\text{T}}]_{\beta\alpha} 
\end{split}
 \label{eqn:suM1-CCR-3}
\\
& [ \mathcal{K} , \mathcal{Q}_{\alpha} ]= - (N - 1)\mathcal{Q}_{\alpha} \, , \quad 
[ \mathcal{K} , \widetilde{\mathcal{Q}}_{\alpha} ]= (N-1) \widetilde{\mathcal{Q}}_{\alpha} \; .
 \label{eqn:suM1-CCR-4}
\end{align}
\end{subequations}
It is helpful to write down the right-hand side of \eqref{eqn:suM1-CCR-2} explicitly in the matrix form:
\begin{equation}
[G^{A}]_{\beta\alpha} \mathcal{S}^{A} + \frac{1}{N} \delta_{\alpha\beta} \mathcal{K} 
= 
\left( 
\begin{array}{c|c}
\mathbf{e}_{\alpha\beta} & 0 \\
\hline
0 &  \delta_{\alpha\beta} 
\end{array}
\right)  \; ,
\label{eqn:suN1-ACR-by-matrices}
\end{equation}
where the $N \times N$ matrices $\mathbf{e}_{\alpha\beta}$ ($\alpha,\beta=1,\ldots,N$) are defined 
by $[\mathbf{e}_{\alpha\beta}]_{ij}= \delta_{i\alpha} \delta_{j\beta}$.   
Out of the above $N(N+2)$ generators, we can construct the quadratic Casimir as:
\begin{equation}
\mathcal{C}_{2} = \sum_{A=1}^{N^{2}-1} \mathcal{S}^{A} \mathcal{S}^{A} 
- \frac{1}{N(N-1)} \mathcal{K}^{2} 
- \sum_{\alpha=1}^{N}  \left( 
\mathcal{Q}_{\alpha} \widetilde{\mathcal{Q}}_{\alpha} 
- \widetilde{\mathcal{Q}}_{\alpha} \mathcal{Q}_{\alpha} 
\right)    \;  .
\label{eqn:quad-Casimir}
\end{equation}
\subsection{Fock representations}
\subsubsection{Abrikosov construction (slave boson)}
There are two different ways to realize the SU($N | 1$) algebra \eqref{eqn:suM1-CCR-1}-\eqref{eqn:suM1-CCR-4} 
in terms of bosons and fermions.  
One is to use $N$ species of (ordinary) fermions $\{ f_{\alpha}^{\dagger} \}$ 
and one species of boson $b$ (construction ``F-B'') which is known 
as the {\em slave-boson representation} \cite{Coleman-84}: 
\begin{subequations}
\begin{align}
& \widehat{\mathcal{S}}^{A} = f_{\alpha}^{\dagger} [S_{A} ]_{\alpha\beta} f_{\beta} \, , \quad 
\widehat{\mathcal{K}} = n_{\text{F}} + N n_{\text{B}}   
\label{eqn:suMN-gen-Abrikosov-1} \\
& \widehat{\mathcal{Q}}_{\alpha} = f^{\dagger}_{\alpha} b \, , \quad 
\widehat{\widetilde{\mathcal{Q}}}_{\alpha} = b^{\dagger}  f_{\alpha}  \quad (\alpha =1,\ldots, N)  
\; ,
\label{eqn:suMN-gen-Abrikosov-2}
\end{align}
\end{subequations}
where the fermion and boson numbers are defined by $n_{\text{F}}= \sum_{\alpha} f^{\dagger}_{\alpha} f^{\alpha}$ 
and $n_{\text{B}} = b^{\dagger}b$, respectively.  
Obviously, $n := n_{\text{F}} + n_{\text{B}}$ is conserved and can be used to label irreducible representations 
of SU($N|1$).  
In fact, the quadratic Casimir in \eqref{eqn:quad-Casimir} is given by:
\begin{equation}
\mathcal{C}_{2}^{\text{F-B}}(n)  = \frac{N}{N-1}  n \left\{ (N - 1) - n\right\}   \;  .
\end{equation}
We can easily check that the choice $n=1$ correctly reproduces the expressions \eqref{eqn:suN1-bosonic-gen} 
and \eqref{eqn:suN1-fermionic-gen}.   
For general $n$, the representation consists of $\text{min}(N,n)+1$ different  irreducible representations 
of SU($N$)
\begin{equation}
\bigoplus_{n_{\text{F}}=0}^{\text{min}(N,n)}
\left[ {\tiny  \text{\scriptsize $n_{\text{F}}$} \left\{ 
\yng(1,1,1,1) \right. } \;\; (n_{\text{B}}= n- n_{\text{F}} ) \right] 
 \label{eqn:SUN1-irreps-FB}
\end{equation}
corresponding to the possible fermion numbers $n_{\text{F}}\,[=0,\ldots, \text{min}(N,n)]$.  

\subsubsection{Schwinger construction (slave fermion)}
Another construction uses $N$ species of bosons $b^{\dagger}_{\alpha}$ and one fermion $f^{\dagger}$ 
(construction ``B-F''; {\em slave-fermion representation}): 
\begin{subequations}
\begin{align}
& \widehat{\mathcal{S}}^{A} = b_{\alpha}^{\dagger} [S_{A} ]_{\alpha\beta} b_{\beta} \, , \quad 
\widehat{\mathcal{K}} = n_{\text{B}} + N n_{\text{F}}   
\label{eqn:suMN-gen-Schwinger-1}   \\
& \widehat{\mathcal{Q}}_{\alpha} = b^{\dagger}_{\alpha} f \, , \quad 
\widehat{\widetilde{\mathcal{Q}}}_{\alpha} = f^{\dagger}  b_{\alpha}  \quad (\alpha =1,\ldots, N)  
\; .
\label{eqn:suMN-gen-Schwinger-2}
\end{align}
\end{subequations}  
The boson and fermion numbers are defined by $n_{\text{B}}= \sum_{\alpha} b^{\dagger}_{\alpha} b^{\alpha}$ 
and $n_{\text{F}} = f^{\dagger}f$, respectively, 
and the quadratic Casimir $\mathcal{C}_{2}$ now is determined by $n= n_{\text{F}} + n_{\text{B}}$ as:
\begin{equation}
\mathcal{C}_{2}^{\text{B-F}}(n)  = \frac{N-2}{N-1} n \left\{ (N-1) + n \right\}  \;  .
\end{equation}
Since the fermion number can take only two values $n_{\text{F}}=0,1$, 
the representation specified by $n$ is made of two SU($N$) irreducible representations:
\begin{equation}
\left[ \underbrace{ {\tiny 
\yng(4) }}_{n} \;\; (n_{\text{F}}= 0) \right] \oplus 
\left[ \underbrace{ {\tiny 
\yng(3) }}_{n-1} \;\; ( n_{\text{F}}= 1) \right] 
 \; .
 \label{eqn:SUN1-irreps-BF}
\end{equation}
Although the conserved $n = n_{\text{F}} + n_{\text{B}}$ again plays a crucial role in specifying the irreducible representations, 
the representations \eqref{eqn:suMN-gen-Abrikosov-1}-\eqref{eqn:suMN-gen-Abrikosov-2} and 
\eqref{eqn:suMN-gen-Schwinger-1}-\eqref{eqn:suMN-gen-Schwinger-2} in general realize {\em different} 
irreducible representations even for the same $n$ (except for $n=1$). 

The $n=2$ representation used in Sec.~\ref{sec:SUSY-FM-JK} is constructed as follows.  
First, the $N(N+3)/2$ basis states are given by:
\begin{subequations}
\begin{align}
\begin{split}
& (n_{\text{B}},n_{\text{F}})=(2,0) :  \ldots {\tiny \yng(2)}  \\
& \qquad 
|(\alpha,\alpha)\rangle = \frac{1}{\sqrt{2}} (b_{\alpha}^{\dagger})^{2} |0\rangle \\
& \qquad 
|(\alpha,\beta) \rangle = b_{\alpha}^{\dagger} b_{\beta}^{\dagger} |0\rangle \quad (\alpha < \beta) \; .
\end{split}
\label{eqn:B-F-n2-rep-basis-1}
 \\
 \begin{split}
& (n_{\text{B}},n_{\text{F}})=(1,1) :  \ldots {\tiny \yng(1)}  \\
& \qquad 
|\boldsymbol{N}; \alpha\rangle = b_{\alpha}^{\dagger} f^{\dagger} |0\rangle \quad (\alpha=1,\ldots, N)   \; .
\end{split}
\label{eqn:B-F-n2-rep-basis-2}
\end{align}
\end{subequations}
We can find all the matrix elements of the generators by applying the expressions 
\eqref{eqn:suMN-gen-Schwinger-1}--\eqref{eqn:suMN-gen-Schwinger-2} to the above basis states.  
%
\end{document}